\documentclass[ALICE,manyauthors]{cernphprep}

\usepackage[comma,square,numbers,sort&compress]{natbib}
\usepackage{hyperref}
 \usepackage{lineno}
\newcommand{\Ngammalike}{$N_{\rm {\gamma-like}}$} 
\newcommand{\Ngammatrue}{$N_{\rm{\gamma-true}}$} 
 
\newcommand{\pp}{\mbox{pp}}
\newcommand{\sqrts}{$\sqrt{s}$}
\newcommand{\etawindow}{$2.3 < \eta <  3.9$}
\newcommand{\aven}{$\langle n \rangle$}
\begin{document}%

%%%%%%%%%%%%%%%  Title page %%%%%%%%%%%%%%%%%%%%%%%%
\begin{titlepage}
\PHyear{2014}
\PHnumber{280}      % required, will be obtained from PH
\PHdate{17 November}  % required, will be obtained from PH
%
 %%% Put your own title + short title here:

  \title{Inclusive photon production at forward rapidities\\in
  proton-proton collisions at $\mathbf{\sqrt{s}}$ = 0.9, 2.76 and 7
  TeV} 
\ShortTitle{Inclusive photon production in pp collisions}  % appears on right page headers

  % 
  %%% Do not change the next lines!

\Collaboration{ALICE Collaboration\thanks{See Appendix~\ref{app:collab} for the list of collaboration members}}
\ShortAuthor{ALICE Collaboration} % appears on left page headers, do not change

  \begin{abstract}	
    The multiplicity and pseudorapidity distributions of inclusive photons
have been measured at forward rapidities (\etawindow) in proton-proton
collisions at three center-of-mass energies, \sqrts~= 0.9, 2.76 and
7~TeV using the ALICE detector.  
It is observed that the
increase in the average photon multiplicity as a function of beam
energy is compatible with both a logarithmic and a power-law dependence.
The relative increase in average photon multiplicity produced in
inelastic pp collisions at 2.76 and 7~TeV center-of-mass energies with
respect to 0.9~TeV are 37.2\% $\pm$ 0.3\% (stat) $\pm$ 8.8\% (sys) and
61.2\% $\pm$ 0.3\% (stat) $\pm$ 7.6\% (sys), respectively. The photon multiplicity
distributions for all center-of-mass energies are well described by
negative binomial distributions. The multiplicity distributions are
also presented in terms of KNO variables. The results are compared to
model predictions, which are found in general to underestimate the
data at large photon multiplicities, in particular at the highest
center-of-mass energy. 
Limiting fragmentation behavior of photons has been explored with the data, but is not
observed in the measured pseudorapidity range.  %%%%%%%%%%% put the abstract
  \end{abstract}

\end{titlepage}
\setcounter{page}{2}
\section{Introduction}

The Large Hadron Collider (LHC) at CERN offers the possibility to
study particle production mechanisms in proton-proton~(\pp) collisions
at unprecedented center-of-mass energies. 
Measurements of multiplicity and pseudorapidity distributions of
produced particles in \pp~collisions are important for the study of
particle production mechanisms and to obtain the baseline distributions 
for heavy-ion collisions. 
Charged particle measurements in \pp~collisions
at the LHC for central rapidities were reported by the
ALICE~\cite{ALICE-pp1,ALICE-pp2}, CMS~\cite{CMS-pp} and ATLAS~\cite{ATLAS-pp} 
collaborations and at forward rapidities by the LHCb collaboration~\cite{LHCb-pp1,LHCb-pp2}.
Inclusive photon measurements provide complementary information to those of charged
particles as the majority of the photons are decay products of neutral pions.  
Measurements at forward rapidities enable an extension of the study
of particle production mechanisms carried out at mid-rapidities. 

In the present work, we report the measurement of inclusive photon
production in the forward pseudorapidity region, \etawindow,
for \pp~collisions at
\sqrts~=~0.9, 2.76 and 7~TeV, with the ALICE detector. 
Multiplicity and spatial distribution of photons are measured on an
event-by-event basis by the
Photon Multiplicity Detector (PMD), which exploits the pre-shower
photon measurement technique. 
We present the beam-energy dependence of the average photon multiplicity and
pseudorapidity distributions of photons.
The pseudorapidity distributions, plotted with respect to the
corresponding beam rapidities, are used to test the predictions of the
limiting fragmentation behavior~\cite{limiting1}.
The results are compared to different tunings of 
PYTHIA~\cite{PYTHIA} and PHOJET~\cite{PHOJET1,PHOJET2} models.

This paper is organized as follows. Section~2 describes the experimental
setup for the measurement of photons using the PMD. 
Event selection and trigger settings are discussed in section~3.
A discussion on the event generators and simulation framework is given
in section~4. Performance of the PMD modules for incident charged
particle and electron beams are discussed in section~5.
Photon reconstruction is presented in
section~6 and the unfolding method used to correct for detector effects is described
in section~7. The study of the systematic uncertainties is outlined in section~8. 
The results of photon multiplicities and pseudorapidity distributions are
discussed in section~9. 
We conclude with a summary and outlook in section~10.

\section{Experimental setup}

The ALICE detector~\cite{ALICE_expt}
consists of a large number of detector subsystems, and
has a unique potential for \pp~physics in terms of 
excellent primary and secondary vertex reconstruction capabilities.
The central barrel consists of the Inner Tracking System (ITS), the
Time Projection Chamber,  the Transition Radiation Detector, 
the Time of Flight detector and the electromagnetic
calorimeter. 
The two Silicon Pixel Detector (SPD) layers of the ITS 
surround the central beryllium beam pipe and cover the pseudorapidity
ranges, $|\eta|<2$ and $|\eta|<1.4$ for the inner and outer layers, respectively.
The central barrel  also includes a High Momentum Particle
Identification Detector and a Photon Spectrometer.
The Muon Spectrometer and the PMD are both located at forward rapidities,
but on opposite sides of the nominal interaction point.
The present analysis
uses data from the PMD for photon
reconstruction. 
Several sets of trigger and multiplicity detectors are placed at
forward rapidities, which include 
the Forward Multiplicity Detector, 
the Zero Degree Calorimeter and detectors for trigger and
timing (V0 and T0). The V0 detector consists of two scintillator
arrays, placed on either side of the interaction region, at 
$2.7<\eta<5.1$ and $-3.7<\eta<-1.7$. It is used for event selection
and background rejection.
The Minimum Bias (MB) trigger conditions are achieved 
by a combination of signals from V0 and SPD.

The PMD~\cite{PMD_TDR,PMD_ATDR} 
is located at a distance of 367~cm from the interaction
point and spans a pseudorapidity region between
2.3 and 3.9, with full azimuthal coverage. 
The PMD makes use of the pre-shower technique where a three radiation
length ($X_{\rm 0}$) thick lead converter is sandwiched between two planes of highly
granular gas proportional counters. 
The granularity and the converter thickness of the PMD are optimized
for high particle density. 
The PMD consists of 184320 honeycomb shaped gas cells arranged in
40 modules in the two planes.  
Each cell is of 0.22 cm$^2$ area, and has a honeycomb shaped cathode
extended towards a 20~$\mu$m thick gold-plated tungsten wire at
ground potential at the center of each cell. 
The sensitive medium is a gas mixture of Ar and CO$_2$ in a 
70:30 ratio. The front-end electronics
consists of MANAS chips for anode signal processing and the Cluster
Read Out Concentrator Unit System (CROCUS) for data
acquisition~\cite{PMD_ATDR}. The PMD is assembled in two equal
halves. Each half has independent cooling, gas supply and electronics
accessories. 
A photon traversing the converter plate produces an
electromagnetic shower in the pre-shower plane, leading to a large
signal, spread over several cells. The signal from a charged particle, on the other
hand, is confined mostly to a single cell. The differences in the
responses of charged particles to photons are used to reject
charged tracks in the analysis.

\section{Event selection}

The data used in this analysis are from \pp~collisions at
center-of-mass energies of 0.9, 2.76 and 7~TeV, collected by the ALICE
detector with a magnetic field of 0.5~T.
The number of analyzed events are 2, 8 and 9~millions at \sqrts~= 0.9, 2.76 and
7~TeV, respectively.
Data were taken in 2010 and 2011 under conditions where pileup 
effects were small. The probability of collision pileup per
triggered event was below 3\%. 
Events with more than one vertex reconstructed with the SPD was
rejected to minimize the effect of pileup. This rejection is
especially effective at rejecting average and high multiplicity
pileup events which can contaminate the multiplicity distribution. 

The interaction vertex is reconstructed using the two silicon pixel
layers of the SPD. 
For the present analysis, we restrict the measured $z$-vertex to
be within $\pm 10$~cm from the nominal interaction point. 
The minimum bias data used for the  inelastic (INEL) events were collected
using a minimum bias trigger
(MB$_{\rm OR}$) condition,
which requires at least one hit in
the SPD or in either of the two V0 arrays~\cite{ALICE-pp1}. 
This condition is satisfied by
the passage of a charged particle anywhere in the 8 units of
pseudorapidity covered by these detectors.
The analysis for non-single diffractive (NSD)
events requires a coincidence between the
two sides of the V0 detectors. Except for the trigger selection criteria,
identical data analyses procedures were followed for both INEL and NSD
events. The experimental results are corrected for trigger and vertex
reconstruction efficiencies, estimated by means of Monte Carlo (MC) simulations~\cite{diffraction}.
Table~\ref{Trigg} lists the two efficiencies.
The systematic errors due to these efficiencies affect the zero
multiplicity bin, but are negligible for other multiplicity bins.
For the zero bins, these systematic errors are 18\%, 29.3\%
and 26.7\% for the data at 0.9, 2.76 and 7~TeV, respectively. 
Multiplicity-dependent correction factors are used to obtain the photon multiplicity
distributions, whereas an overall correction factor is applied 
for pseudorapidity densities of photons.
\begin{table}[ht]
\begin{center}
\begin{tabular}{|c |c |c|}
\hline
$\sqrt{s}$ (TeV)   & MB$_{\rm OR}$ Trigger &  Vertex
reconstruction  \\
           & efficiency (\%)&  efficiency (\%)  \\
\hline
0.9  & $91.0{^{+3.2}_{-1.0}}$ & 91.4 $\pm$ 1.7\\
\hline
2.76  & $88.1{^{+5.9}_{-3.5}}$ & 92.4 $\pm$ 1.1\\
\hline
7.0  & $85.2{^{+6.2}_{-3.0}}$ & 92.8  $\pm$ 1.2\\ 
\hline
\end{tabular}
\caption{The minimum bias trigger efficiency and the vertex
  reconstruction efficiency for pp collisions.}
\label{Trigg}
\end{center}
\end{table}

\section{Event generators and simulation framework}

Corrections for the instrumental effects in the photon measurement and
the estimation of systematic uncertainties are performed based on
simulated events using various tunes of  PYTHIA and PHOJET event generators.
The PYTHIA event generator combines perturbative QCD and 
phenomenologically motivated models.
PYTHIA employs several tunable parameters, which result in different
tunes of the event generator, such as 
PYTHIA 6.4~\cite{PYTHIA} tune D6T~\cite{PYTHIAD6T},
Perugia-0~\cite{PYTHIAPerugia0}, and
ATLAS-CSC~\cite{csc}). 
The PHOJET~1.2 generator~\cite{PHOJET1,PHOJET2} is based on a 
two-component approach, that describes 
high-energy collisions as an interplay of soft and hard components. 

The response of the PMD to the produced particles
is studied using PYTHIA and PHOJET event generators.
The AliRoot~\cite{AliRoot} software package, which includes a detailed
information of ALICE apparatus, has been utilized.
Particle transport of the generated particles from the event
generators have been simulated using the GEANT-3~\cite{GEANT3} software
package, and stored in terms of energy depositions and positions of
the hits. The deposited energy is converted to ADC values using the
energy loss conversion relation to treat the simulated data on a 
similar footing as the experimental data. The MC description of the
cluster sizes has been validated using test beam data. The
distribution of cluster sizes for photon candidates was found to
reproduce well the measured distribution, even in the tail region.

\section{Performance of the PMD}

The performance of the PMD has been studied by 
exposing the detector modules to pion and
electron beams at energies ranging from 
1~to~6~GeV at the CERN Proton Synchrotron.
Two modules were mounted back to back on a movable stand. A gap was
maintained between the modules to place different thickness of lead
converters for the pre-shower study. Data were readout by using same
front-end electronics and data acquisition as in the real
experiment. Special trigger combinations, using scintillator paddles and Cherenkov
counters, were configured for pion and electron beams separately. Data
were collected for various combinations of beam energy, converter
thickness and detector operating voltage. 
The responses of the modules to pion and electron beams have been
simulated by using GEANT-3 code, to understand the
observed data.

\begin{figure}[tbh!]
\begin{center}
\includegraphics[width=0.4\textwidth]{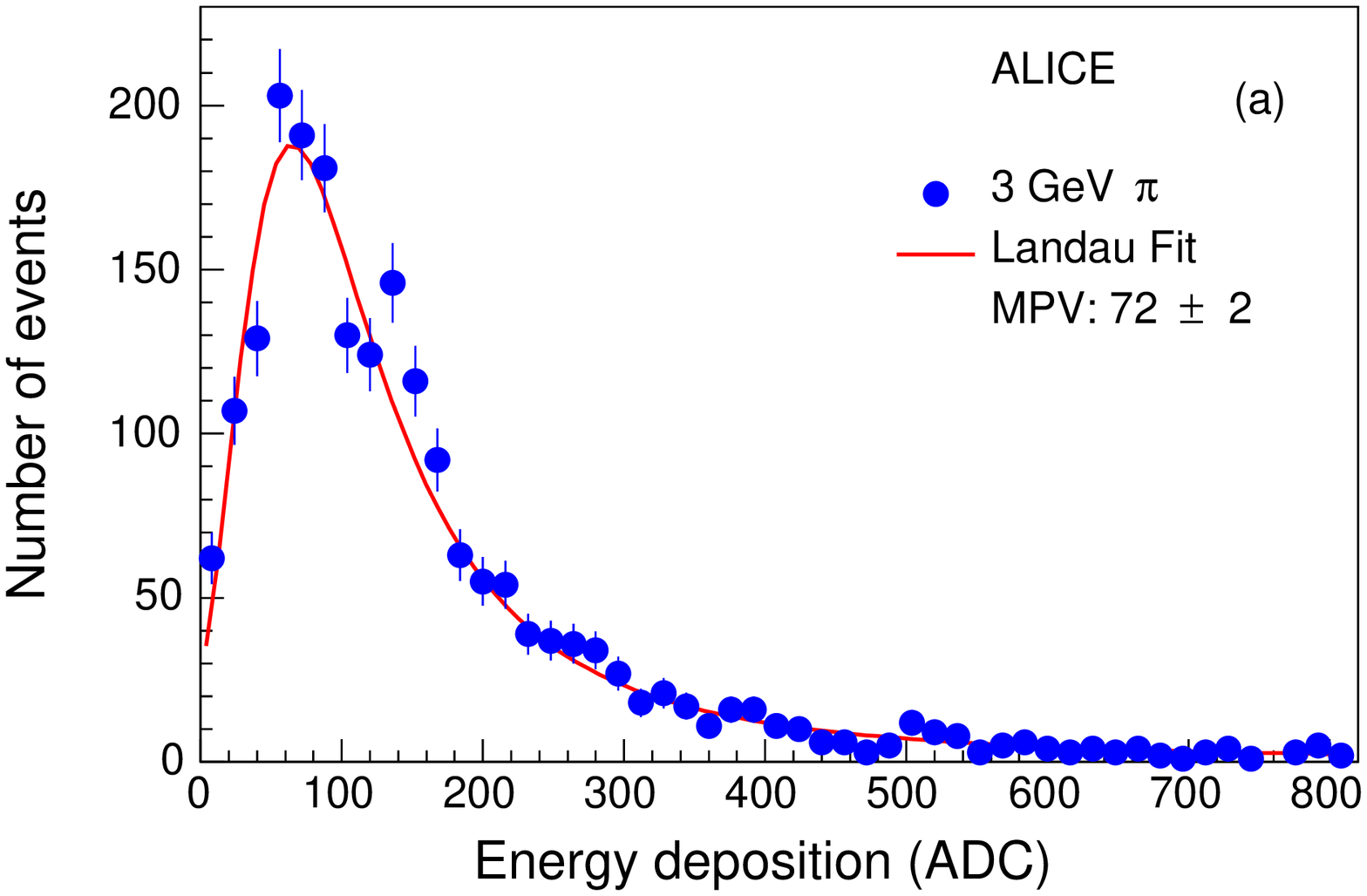}
\includegraphics[width=0.4\textwidth]{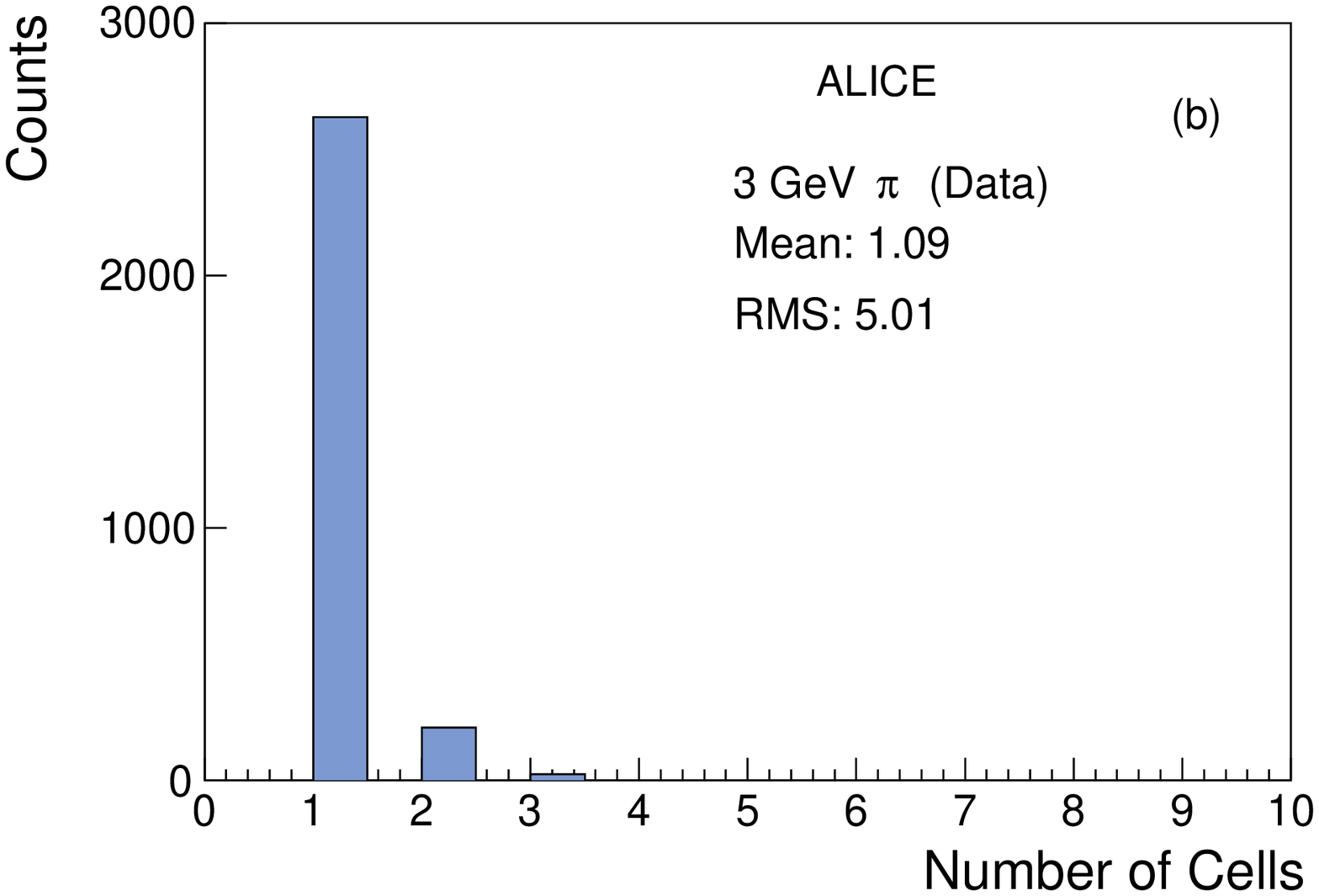}
\includegraphics[width=0.4\textwidth]{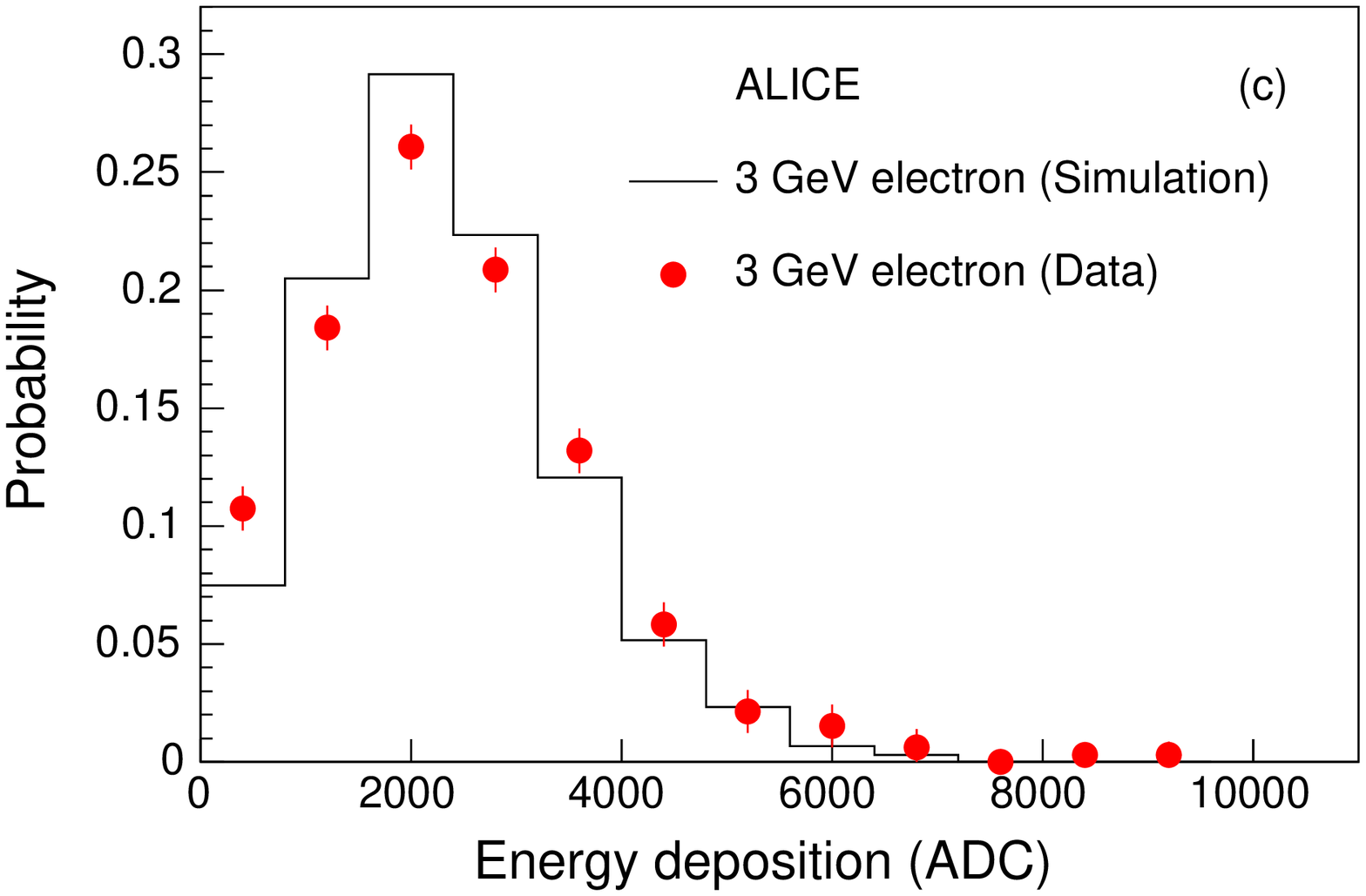}
\includegraphics[width=0.4\textwidth]{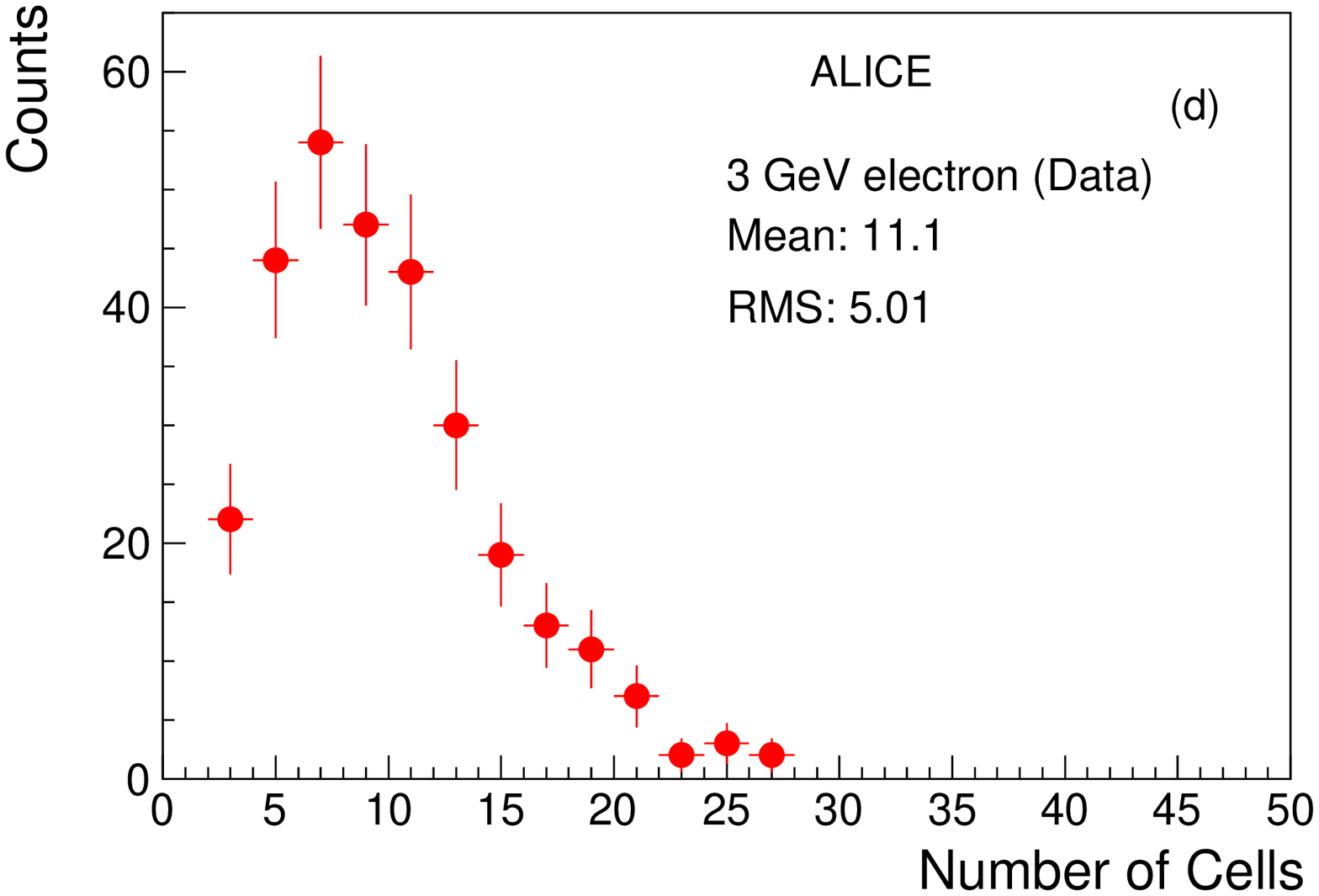}
\caption{(Color online). 
Panels (a) and (c) show the
energy depositions in the PMD module for 3~GeV pions and 3~GeV
electrons, respectively. The pion distribution is fitted with a Landau fit, which gives
the most probable value (MPV) of energy depositions by 
charged particles. For electrons, the result from the simulation is
superimposed on the experimental data. Panels (b) and (d) show the
number of cells hit for 3~GeV pions and 3~GeV electrons, respectively.  
Note the large difference in scales in the abscissa for pions and electrons.
}
\label{perform}
\end{center}
\end{figure}

Results for the performance of the PMD modules are presented in
Fig.~\ref{perform} for an operating voltage of $-1300$~V. 
Energy deposition (in terms of ADC) and number of cells hit for 3 GeV
pions are shown in the top panels of the figure.
The most probable value (MPV) of the energy deposition by charged
particles is obtained by fitting to a Landau distribution function,
which yields an ADC value of $72\pm 2$. 
This MPV value is used later
in the manuscript to
discriminate charged particle contamination in the photon sample. 
Pre-shower characteristics were studied by placing the lead converter
and by bombarding the detector with electron beams of different
energies. The bottom panels of the figure show the response of the
module to 3~GeV electron beams for a 3$X_{\rm 0}$ converter thickness.
The mean energy deposition of $\sim$2000 ADC is much larger than
that of the response of the pions. It has also been
observed that the signals from a charged hadrons affect on average $\sim$1.1
cells, whereas signals from 3~GeV electrons affect 11~cells.
Thus, compared to the charged pions, the electrons deposit 
larger amounts of energies and affect larger number of cells.

\begin{figure}[tbh!]
\begin{center}
\includegraphics[width=0.5\textwidth,height=0.25\textheight]{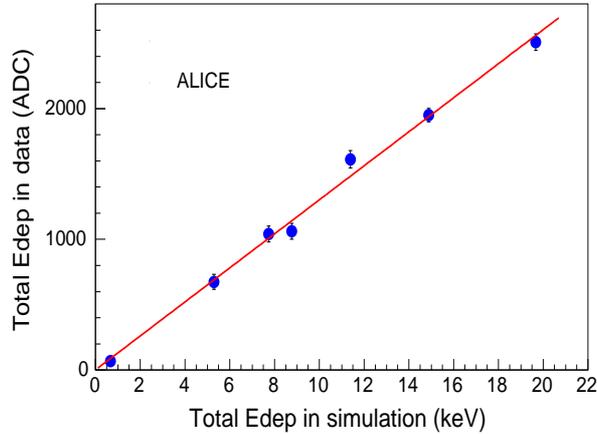}
\caption{(Color online). 
Relationship between the mean energy depositions in the PMD modules
obtained from simulated results
(in keV) and experimental data (in ADC)
for pion and electron beams of various energies. 
}
\label{perform1}
\end{center}
\end{figure}

Mean energy depositions in the detector modules have been obtained
from both experimental data and simulations.
Fig~\ref{perform1} shows the relationship between the 
mean energy depositions, obtained from the simulated data (in keV) and
the experimental data (in ADC) for pion and electron beams of
different energies. 
A linear fit gives the conversion relation from the simulated to experimental data.
This relation has been used in the analysis chain 
to convert the keV scale of the simulated data to total ADC of the
experimental data.

During the data taking period at the LHC, all the modules of PMD are 
operated at $-$1300~V, where the efficiency for charged pions is
$\sim$90\%. Responses of the cells to minimum bias particles have been
obtained by storing single isolated clusters from the pp collisions at
7~TeV. Comparison of the MPV values for the cells give the 
cell-to-cell gain variation. For the full detector, the cell
to cell gain variation is observed to be within 9\%.

\section{Photon reconstruction}

Particles entering the pre-shower plane of the PMD are expected to affect
more than one cell in general. To analyse the data, the cell hits were
first clustered using a nearest neighbor clustering algorithm.
Thus a cluster is formed by a number of contiguous cells having
non-zero energy deposition.
For each event, total number of clusters are obtained with
corresponding cluster parameters, such as, 
number of hit cells in a cluster, position of the centroid of each cluster and the total 
energy deposition of the cluster. 

\begin{figure}[tbh!]
\begin{center}
\includegraphics[width=0.5\textwidth,height=0.27\textheight]{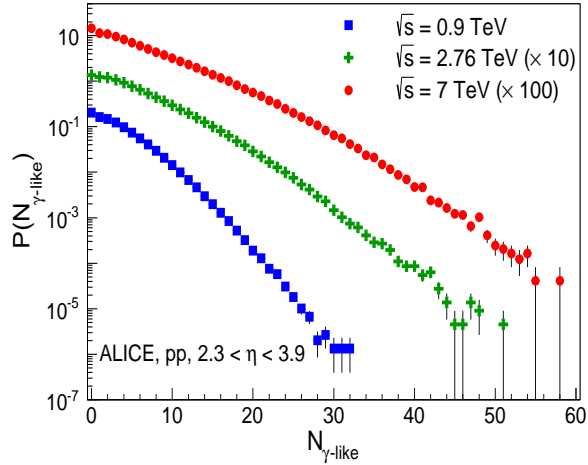}
\caption{(Color online). 
Probability distribution of the
measured  \Ngammalike~clusters (number of clusters above the
photon-hadron discrimination thresholds)
in \pp~collisions at
\sqrts~=~0.9, 2.76 and 7~TeV. 
}
\label{NGammalike}
\end{center}
\end{figure}

To enrich the photon samples in the data, suitable 
photon-hadron discrimination thresholds on the number of 
hit cells and on the energy deposited in clusters, have been applied.
The number of clusters in an
event which pass through the threshold cuts are labelled as 
photon-like clusters, \Ngammalike.
Figure~\ref{NGammalike} shows the
distributions of the number of \Ngammalike~clusters for the three energies
within the detector acceptance of \etawindow, obtained with the
application of discrimination thresholds,
the number of cells greater than 2 and
energy deposition greater than 9 times the MPV. 

\section{Correction for detector effects: method of unfolding}

The material budget affects the photons originating at the vertex and
travelling to the PMD. Photons and charged particles suffer from
rescattering effects in upstream material and may
interact with the detector material to produce secondary particles.
%split to two or more secondary particles. 
The \Ngammalike~clusters are photon-rich clusters with a
contamination of charged particles and secondaries which pass
the threshold cut. 
The effects associated with finite detector acceptance and 
finite efficiency needs to be taken into account in the photon counting.
These effects are handled by using detailed simulation
and applying unfolding procedures.

The detector effects are modeled using a response
function matrix ($R_{\rm mt}$). The matrix elements $R_{\rm mt}$ represent the conditional
probability of measuring a true multiplicity $t$ as a measured
multiplicity $m$. The measured distribution, $M$, can thus be expressed
as the product of response matrix and the true distribution $T$,
\begin{eqnarray}
  M = R_{\rm mt}  T.
\end{eqnarray}
One can therefore obtain the true distribution $T$ for given $M$:
\begin{eqnarray}
T = R_{\rm mt}^{-1} M.
\end{eqnarray}
However the matrix, $R$, may be singular and can not always be inverted
analytically. Furthermore, even if the exact solution exists, 
it oscillates due to finite statistics in the measured distribution. A
regularized unfolding method based on $\chi^2$~minimization is used to
overcome this problem~\cite{Unfolding1,  ALICE-pp1, jan-fiete}. 

The unfolded multiplicity
distribution, $U(N_\gamma)$, is found by minimizing the $\chi^2$ function, which is
defined as: 
\begin{eqnarray}
\hat{\chi}^2(U) = \sum_m \Big({\frac{M_m - \sum_t R_{mt}U_t }{ e_m}}\Big)^2 + \beta P(u),
\end{eqnarray} 
where $e_m$ is the estimated measurement error, $P(u)$ is the
regularization term and $\beta$ is the regularization coefficient. 
$\beta P(u)$ suppresses high frequency components in the solution. 

\begin{figure}[tbh!]
\begin{center}
\includegraphics[width=0.5\textwidth]{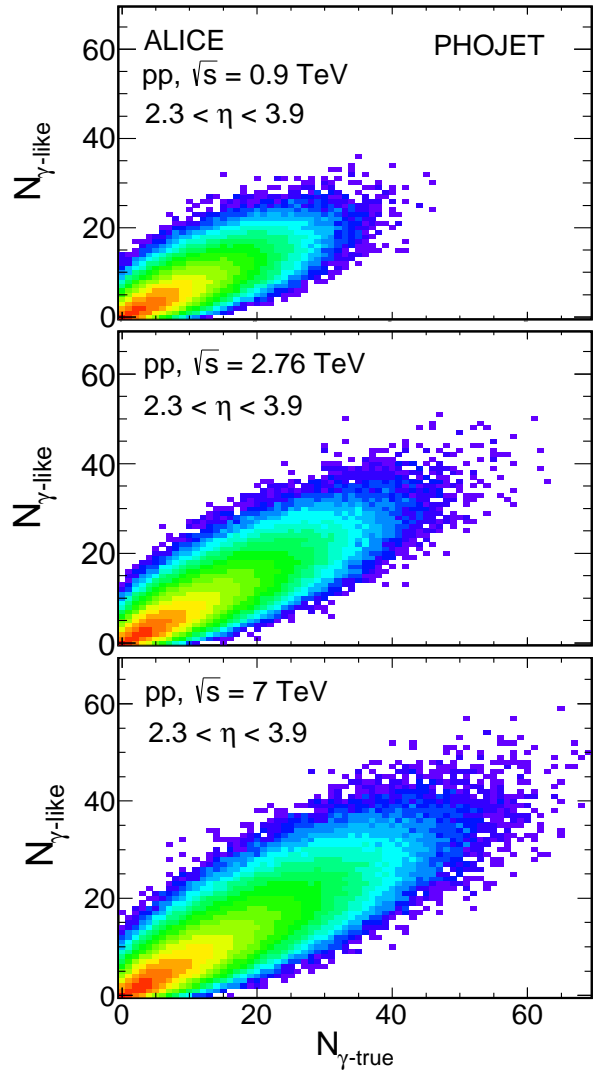}
\caption{(Color online). Detector response matrices in 
\pp~collisions at \sqrts~=~0.9, 2.76 and 7 TeV
 using PHOJET event generator. \Ngammalike~and \Ngammatrue~denote the
 measured and true photon multiplicities,  respectively. 
}
\label{ResponseMatrix}
\end{center}
\end{figure}

\begin{figure}[tbp]
\begin{center}
\includegraphics[width=0.27\textheight]{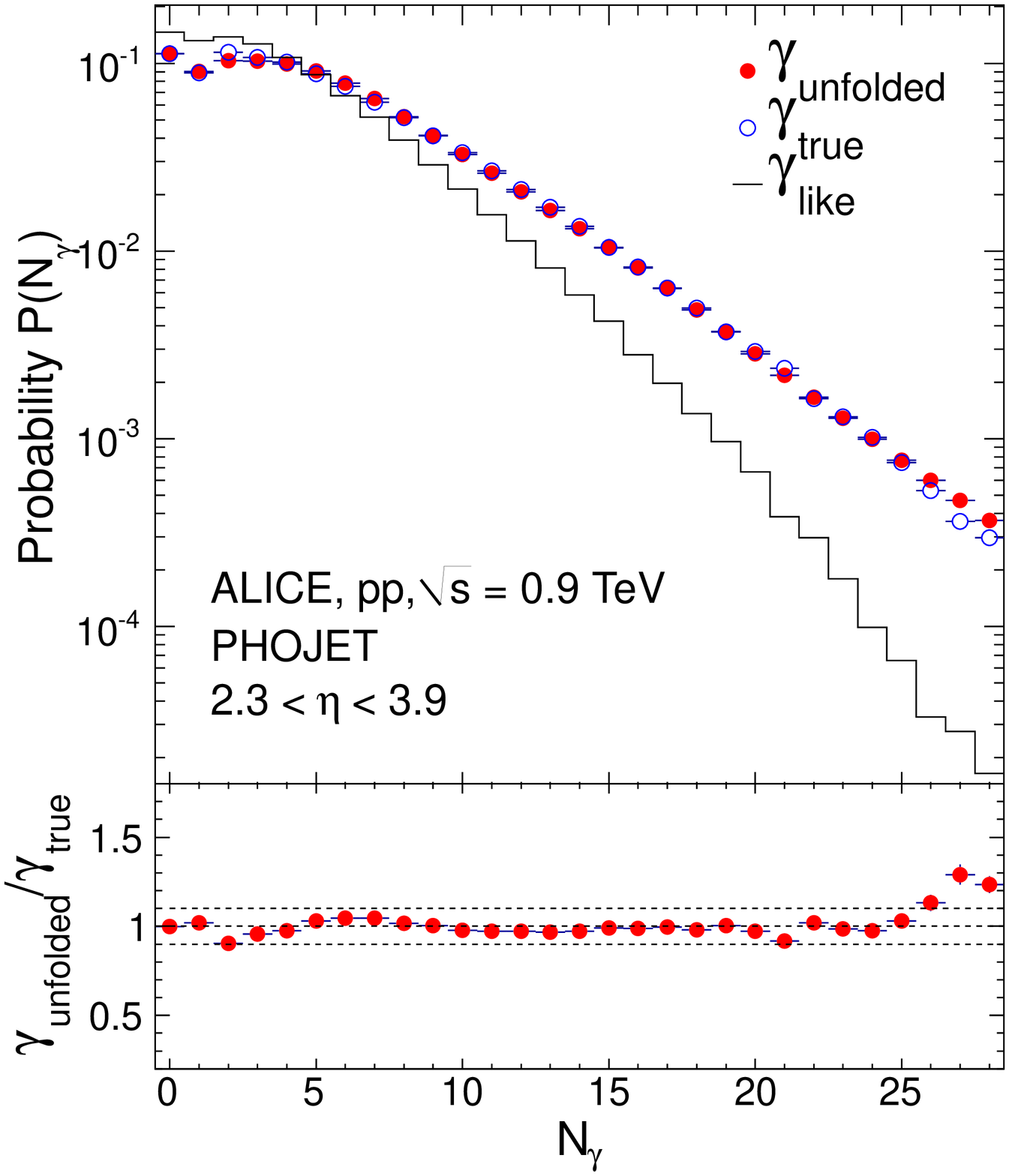}\\
\includegraphics[width=0.27\textheight]{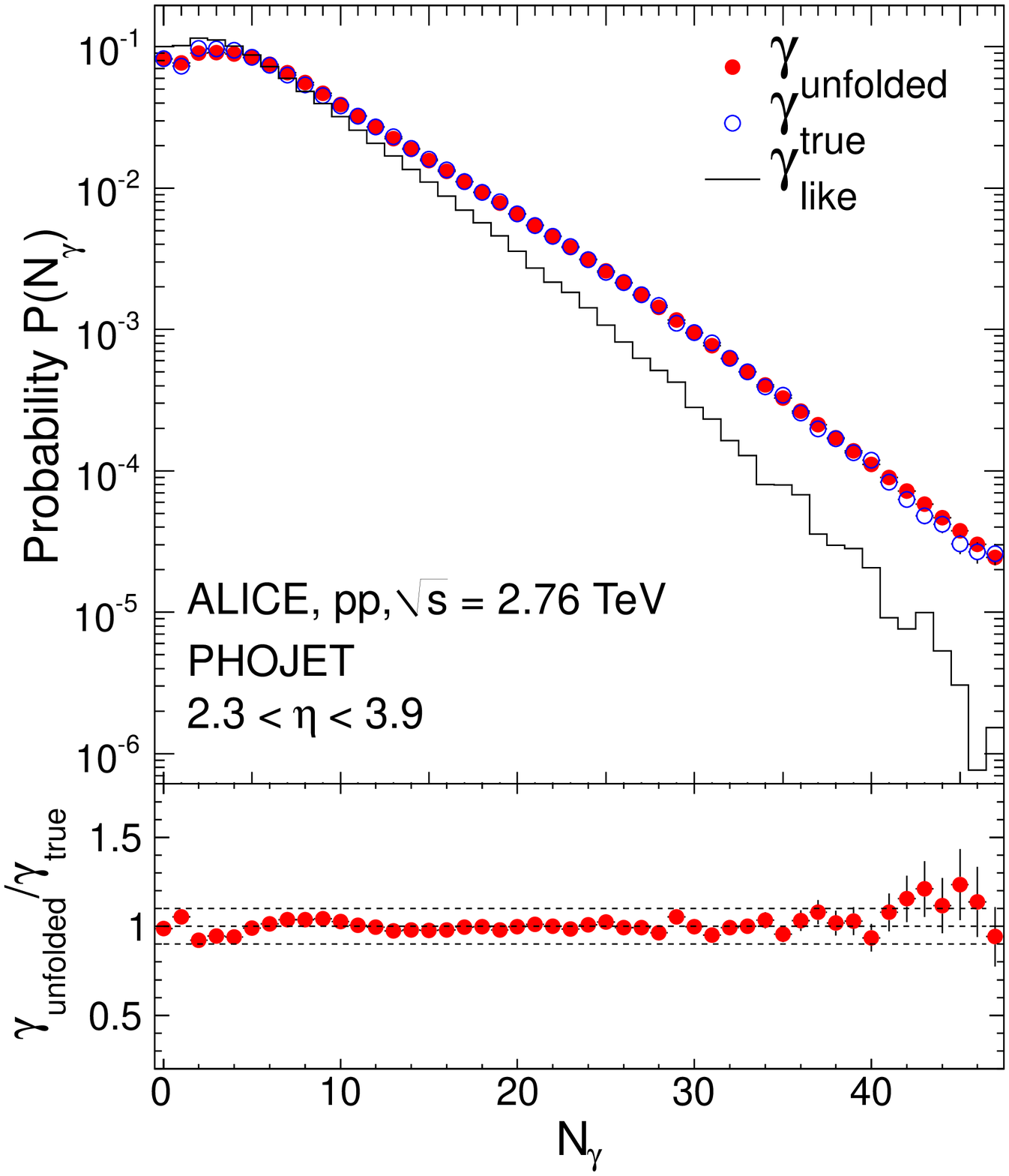}\\
\includegraphics[width=0.27\textheight]{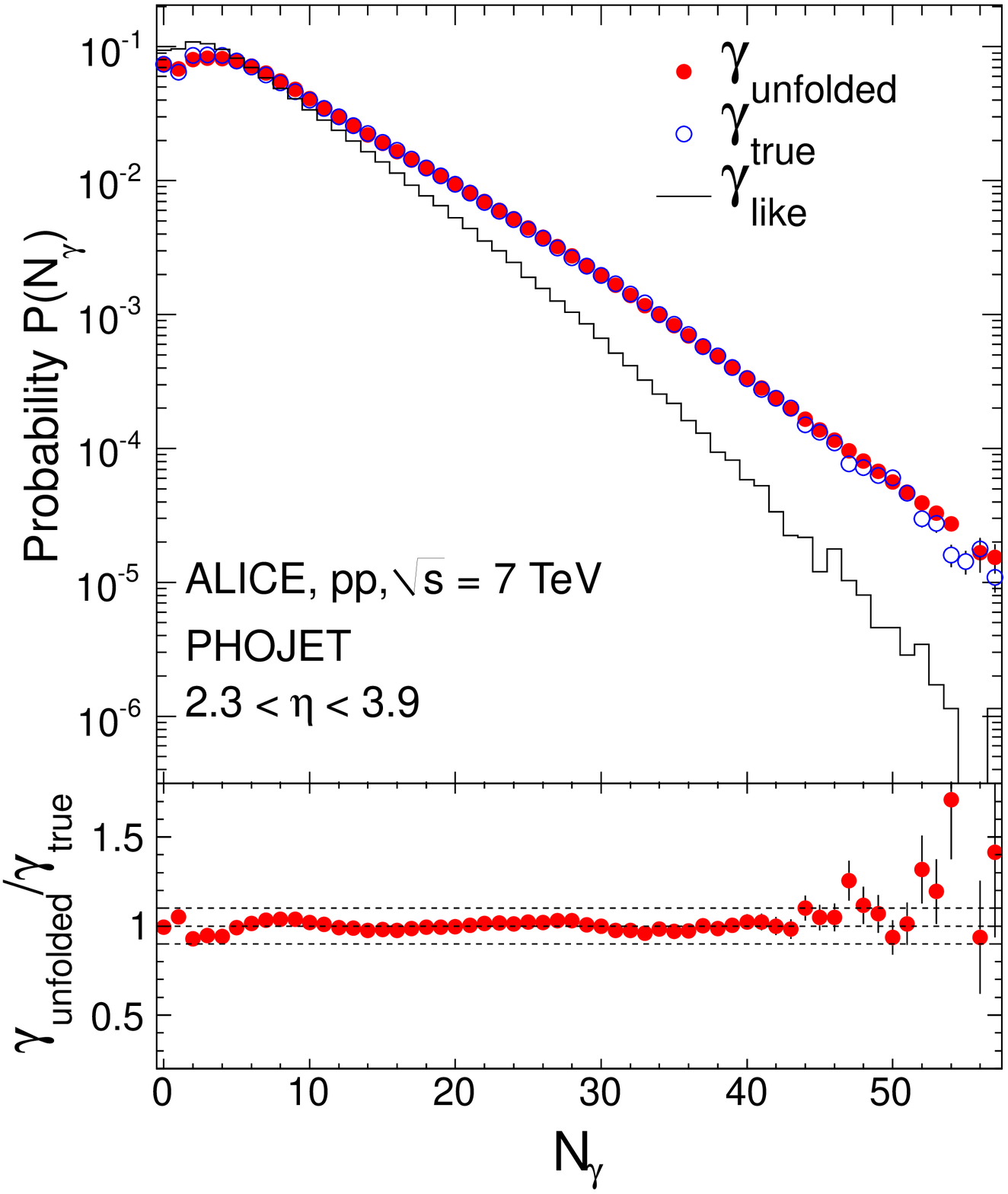}
\caption{(Color online). Test of the unfolding method using PHOJET event
generator for  \pp~collisions at \sqrts~=~0.9 (top), 2.76 (middle)
and 7~TeV (bottom). The measured, unfolded and true photon multiplicities are presented.
The lower panels show the ratios of unfolded to true multiplicity distributions.
}
\label{SimulationTest}
\end{center}
\end{figure}

Figure~\ref{ResponseMatrix} shows the response matrices, 
constructed using the PHOJET event generator within 
\etawindow~in \pp~collisions at $\sqrt{s}$ = 0.9, 2.76 and 7~TeV. 
They represent the correlation between the 
true photon multiplicity (\Ngammatrue) of incident photons and 
the measured photon multiplicity (\Ngammalike) obtained after
applying the photon-hadron discrimination criteria 
on the reconstructed clusters.
It was found in MC studies that to unfold the full multiplicity
distribution with good precision two different $\beta$ values were
needed, one for the two
lowest multiplicity (0 and 1) bins and the second one for all other multiplicities.
The performance of the unfolding method has been tested using
simulated data of \pp~collisions at $\sqrt{s}$ = 0.9, 2.76 and 7 TeV as
shown in Fig.~\ref{SimulationTest}. 
The ratios of the unfolded to true multiplicity distributions agree within  10\% except for the highest multiplicities.
Sensitivity of the results were verified by using different
regularization functions, 
with various $\beta$ values. 

An alternate method of unfolding based on 
Bayes' theorem~\cite{bayes} has also been used, which describes the
definite relationship between the probability of an event with $m$
measured particles conditional on another event with $t$ true
multiplicity. This results in:
\begin{eqnarray}
\bar{R}_{tm} = \frac{R_{mt}P_{t}}{\sum_{t^{'}} R_{mt^{'}} P_{t^{'}}},
\end{eqnarray}
where $R$ is the response matrix and $P_t$ is the apriori distribution
of the true spectrum. After obtaining
$\bar{R}_{tm}$, the unfolded distribution~($U$) can be obtained as: 
\begin{eqnarray}
U_{t} = \sum_{m} \bar{R}_{tm} M_{m}.
\end{eqnarray}
The resultant $U_{t}$ of an iteration is used as the new \textit{a priori} 
distribution for the next iteration.

After unfolding, the results are corrected for trigger and vertex
reconstruction efficiencies to obtain the final results.
The above procedure is used to obtain the photon multiplicity. To obtain the pseudorapidity
distribution of photons, the unfolding method has been employed 
separately in eight $\eta$ bins of width 0.2.

\section{Systematic uncertainties}

The systematic uncertainties on the photon multiplicity have 
contributions from several effects. 
  A major source for the systematic uncertainty to photon
  counting comes from the uncertainty in the implementation of 
  all the known material between the vertex and the PMD in AliRoot. 
  The material budget at forward rapidities was intensely studied in
  connection with the charged particle multiplicity measurements in 
  Pb-Pb collisions~\cite{dndeta_fwd}.  Special sub-sample of events,
  where the vertex was displaced between 
  $-$187.5~cm to 375~cm, so that the material budget in front of
  the FMD and V0 detectors was almost negligible, was used to
  benchmark the MC description of the detector material.
  Discrepancies of the order of 6\% were found in the same rapidity
  region covered by the PMD. 
  Based on this study, a conservative uncertainty on the
  material budget of 10\% has been used in the analysis.
 Events with default material description and with the 10\% increase
  in material have been simulated.  
  The response matrix, using the PHOJET event generator, is obtained from the
  default material setting and the unfolding procedure is followed for
  both the default material setting and with 10\% increase of material. The difference
  between the two unfolded multiplicities is quoted as the systematic
  uncertainty to the multiplicity distribution of photons. 
  The same procedure has been adopted for each $\eta$ bin to obtain the
  systematic uncertainties for pseudorapidity distributions of photons. 

The sensitiveness of the PMD to incident photons of different
momenta depends primarily on the conversion efficiency of photons 
in the detector and then on the discrimination criteria. Simulation
studies have shown that the detector is sensitive to 
transverse momenta as low as $\sim$50~MeV. 
The photon multiplicity is extracted after the unfolding method, where
incident photons of all energy are considered, which makes the present
photon measurement inclusive. Systematic uncertainties arising because
of the choice of discrimination threshold and event generators need to
be considered for the photon counting.

The photon-hadron discrimination conditions in terms of 
number of cells hit and energy depositions
were optimized to minimize the contamination from charged particles
and secondaries. The purity of the photon sample with the default discrimination
condition (number of cells greater than 2 and
energy deposition greater than 9 times the MPV) is 65\%, which is
consistent with the design value~\cite{PMD_TDR,PMD_ATDR}.
The uncertainty due to the choice of
the thresholds have been obtained by changing the
discrimination conditions, that is, the
 number of cells greater than 2 and energy depositions
greater than 6 times MPV. In this condition, the purity of the photon
sample reduces to 60\%. 
Photon counting is also affected by the 
non-uniformity in the PMD, which is taken care of by obtaining
the cell-to-cell gain variation. Its effects on the photon multiplicity and pseudorapidity
distributions have been included as a part
of the systematic uncertainty.

The choice of the event generator may affect the photon counting.
Two different generators, PHOJET and
PYTHIA (tune D6T), have been used as a source of systematic uncertainty. 
The choice of the unfolding method may also add to the uncertainty.
This is studied by using two different unfolding methods, 
$\chi^{2}$ minimization and Bayesian methods. 
In addition, different regularization functions during the unfolding
procedure also add to the  systematic uncertainty.

\begin{table}[ht]
\begin{center}
\begin{tabular}{|c|c|c|c|}
\hline
Sources   & 0.9~TeV  &  2.76 ~TeV  & 7~TeV \\
              & (\%) & (\%) &  (\%) \\
\hline
Material effect & 3 - 18 & 3 - 18 & 3 - 18 \\
\hline
Discrimination  & 0.4 - 4.1 & 4.3 - 6.7 & 0.1 - 1.6\\
thresholds  & & & \\
\hline
Event  & 7.5 - 14.6  & 8.9 - 27.8 & 7.8 - 24.7\\
generators  & & & \\
\hline
Unfolding & 3.6 - 6.2 & 5.7 - 6.4 & 5.4 - 6.5\\
methods & & & \\
\hline
Regularization & 4.1 - 9.5 & 1.9 - 9.6 & 3.0 - 10.7 \\
functions  &  &  &  \\
\hline
Cell-to-cell & 2.7 & 2.7 & 2.7 \\
gain variation  &  &  &  \\
\hline
Total  & 10.1 - 26.4 & 12.4 - 35.8 & 10.8 - 33.2\\         
\hline
\end{tabular}
\caption{Systematic uncertainties in the photon multiplicity 
for INEL collisions at three center-of-mass energies. 
The range of the errors corresponds to lowest and highest 
multiplicity values for each energy.}
\label{Table_sys_mult}
\end{center}
\end{table}

\begin{table}[ht]
\begin{center}
\begin{tabular}{|c|c|c|c|}
\hline
Source   &   0.9~TeV  &   2.76~TeV   & 7~TeV \\ 
               &   (\%)  &  (\%)  & (\%) \\ 
\hline
Material effect &  7 & 7 & 7 \\
\hline
Discrimination & 1 - 2 & 1 - 2  & 1 - 2  \\
thresholds & & &\\
\hline
Event  & 1 - 3 & 1 - 1.6 & 3 - 5 \\
generators & & & \\
\hline
Unfolding  & 0 - 0.7  & 0 - 0.8   & 0 - 0.9 \\
methods & & & \\
\hline
Regularization & 0 - 0.6  & 0 - 0.8   & 0 - 0.4 \\
functions  &  &  &  \\
\hline
Cell-to-cell & 0.3  & 0.3   & 0.3 \\
gain variations  &  &  &  \\
\hline
Total & 7 - 7.9  & 7 - 7.5 & 8 - 8.6 \\
\hline
\end{tabular}
\caption{The magnitude of the sources of systematic uncertainties and their
  contributions to the pseudorapidity distributions of photons for INEL
  collisions at each energy,
  spanning the range, \etawindow.
}
\label{Table_sys_dndeta}
\end{center}
\end{table}

Separate response matrices were generated to evaluate the systematic
uncertainties for each of the sources, e.g., for PYTHIA and PHOJET.
Although the contributions from 
these sources are calculated separately, most of the uncertainties are
likely to be correlated. The contributions from different
sources of systematic uncertainties are quoted for photon
multiplicity and pseudorapidity distributions of the photons in 
Table~\ref{Table_sys_mult} and Table~\ref{Table_sys_dndeta},
respectively.  The total systematic uncertainties are obtained by
adding in quadrature systematic uncertainties from the various
sources.

\section{Results and discussions}

In this section, we present multiplicity and 
pseudorapidity distributions of photons within \etawindow. 
The results are compared to 
predictions from PHOJET and various tunes of 
PYTHIA.

\subsection{Multiplicity distributions} 

\begin{figure}[tbh!]
\begin{center}
\includegraphics[width=0.28\textheight]{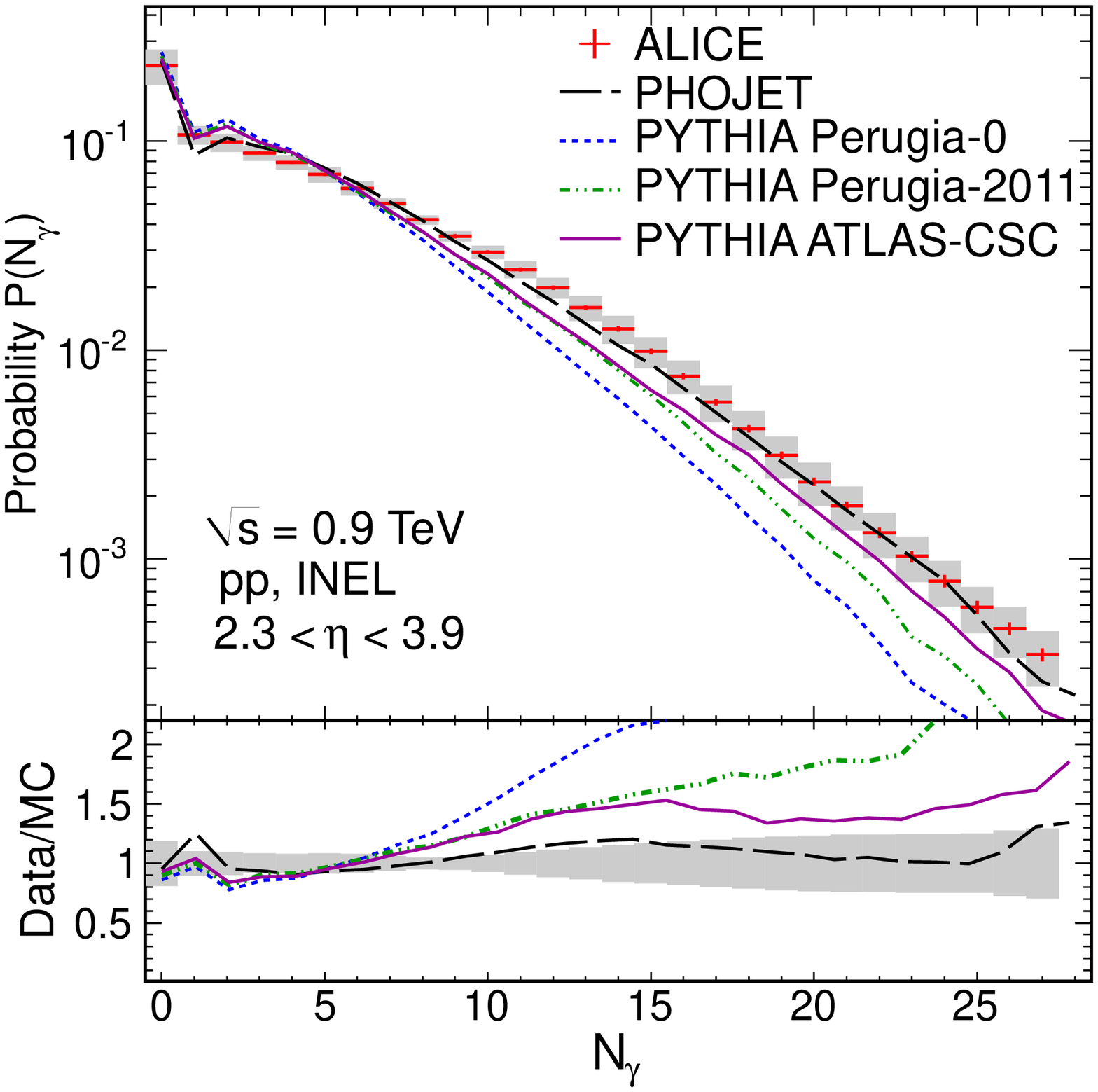}\\
\includegraphics[width=0.28\textheight]{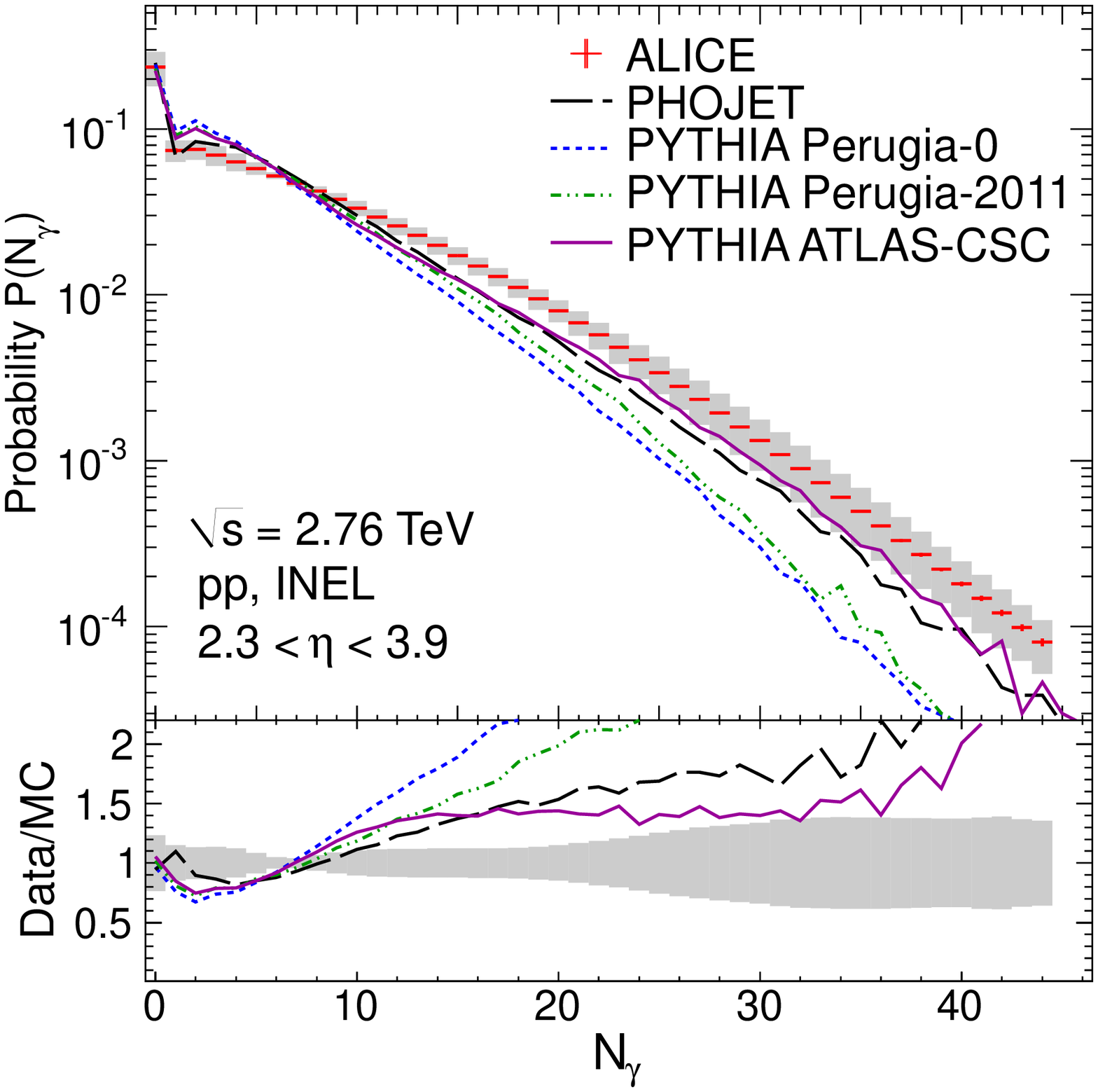}\\
\includegraphics[width=0.28\textheight]{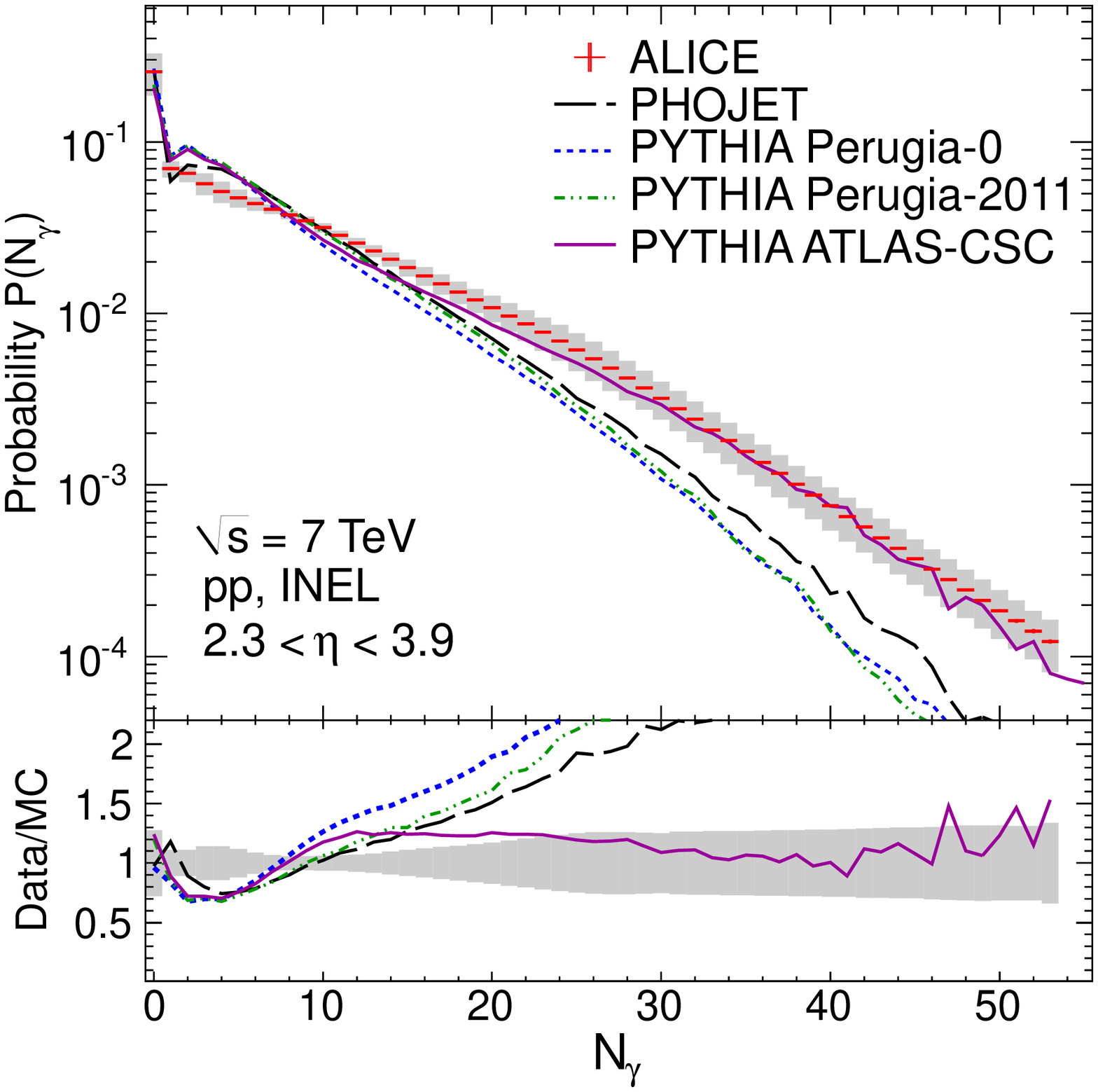}
\caption{(Color online). Multiplicity distributions of photons within
  \etawindow~for 
inelastic events 
in \pp~collisions at \sqrts~=~0.9 (top), 2.76 (middle) 
and 7~TeV (bottom). Predictions from different
event generators are superimposed. Lower panels 
show the ratios between the data and MC distributions. 
The error bars are statistical 
and the shaded regions represent the systematic 
uncertainties. 
}
\label{Mult}
\end{center}
\end{figure} 

Figure~\ref{Mult} shows the multiplicity distributions of photons
for inelastic \pp~collisions within \etawindow~at \sqrts~=~0.9 (top),
2.76 (middle) and 7~TeV (bottom).
The statistical and systematic uncertainties of the data points are
shown by the error bars and shaded bands, respectively.
The average photon multiplicities for the collisions at 0.9 TeV are
$4.46 \pm 0.02$ (stat) $\pm 0.20$ (sys), for 2.76 TeV collisions are 
$6.12 \pm 0.01$ (stat) $\pm 0.47$ (sys), and for 7 TeV collisions are
$7.19 \pm 0.01$ (stat) $\pm 0.45$ (sys). The
relative increase in average photon
multiplicity for 2.76 and 7~TeV 
energies with respect to that of the 0.9~TeV
are 37.2\% $\pm$ 0.3\% (stat) $\pm$ 8.8\% (sys) and
61.2\% $\pm$ 0.3\% (stat) $\pm$ 7.6\% (sys) for inelastic
collisions. 

The multiplicity distributions are compared to the photon multiplicities
obtained from PHOJET, PYTHIA Perugia-0, PYTHIA Perugia-2011, and
PYTHIA ATLAS-CSC. 
The bottom panel for each collision energy shows 
the ratios of experimental data to the event generators.  
PHOJET explains the data at 0.9~TeV, but overestimates the data at all other
energies. Overall, the ATLAS-CSC tune of PYTHIA 
explains the data better at all the energies compared to  
other event generators.
This is consistent with the charged particle multiplicity data at central
rapidities~\cite{ALICE-pp1,ALICE-pp2}.

\begin{figure}[tbh!]
\begin{center}
\includegraphics[width=0.5\textwidth,height=0.28\textheight]{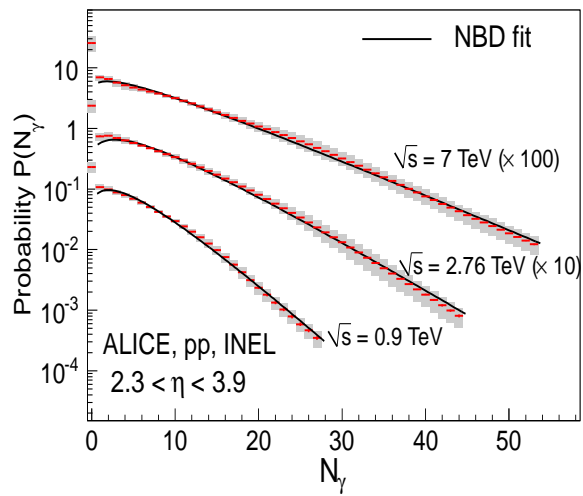}
\caption{(Color online). Multiplicity distribution of photons for
  inelastic collisions, fitted to single NBD functions 
for \pp~collisions at \sqrts~=~0.9, 2.76 and 7~TeV. 
The error bars are statistical 
and the shaded regions represent the systematic 
uncertainties. 
}
\label{NBD}
\end{center}
\end{figure} 

Multiplicity distributions have traditionally been fitted by 
Negative-Binomial-Distributions (NBD) to extract information regarding
the nature of the particle production mechanism~\cite{NBD1,NBD2}. 
The photon multiplicity distributions are fitted with a NBD
function of the form:
\begin{eqnarray}
P(m,k;n) = \frac{\Gamma(n+k)}{\Gamma(n+1)\Gamma(k)}
\frac{(m/k)^{n}}{(m/k+1)^{n+k}},
\end{eqnarray}
where $n$ is the photon multiplicity, 
$m = \langle n \rangle $, $k$ is a parameter responsible for shape of the
distribution.
The data, fitted with NBD function are shown in Fig.~\ref{NBD}.
The fitting is performed for $N_\gamma > 0$.
Fits with the sum of two NBDs (not shown)
do not significantly improve the description of the data.
The parameters of the single NBD fit functions are reported in
Table~\ref{Table_NBD1}. 
With the increase of beam energy,
the average photon multiplicity \aven~increases, whereas
the values of $k$, related to dispersion of multiplicity, decrease.
\begin{table}[ht]
\begin{center}
\begin{tabular}{|c |c |c |}
\hline
$\sqrt{s}$ in TeV & $k$ &  \aven  \\
 \hline
0.9 & 1.89 $\pm$ 0.11 &  5.39 $\pm$ 0.14  \\
\hline
2.76 & 1.72 $\pm$ 0.10  &  7.73 $\pm$ 0.22  \\
\hline
7 & 1.35 $\pm$ 0.07  &  9.03 $\pm$ 0.24  \\
\hline
\end{tabular}
\caption{Fit parameters of single NBD fitting to the photon
 multiplicity distribution.}
\label{Table_NBD1}
\end{center}
\end{table}

\begin{figure}[tbh!]
\begin{center}
\includegraphics[width=0.5\textwidth]{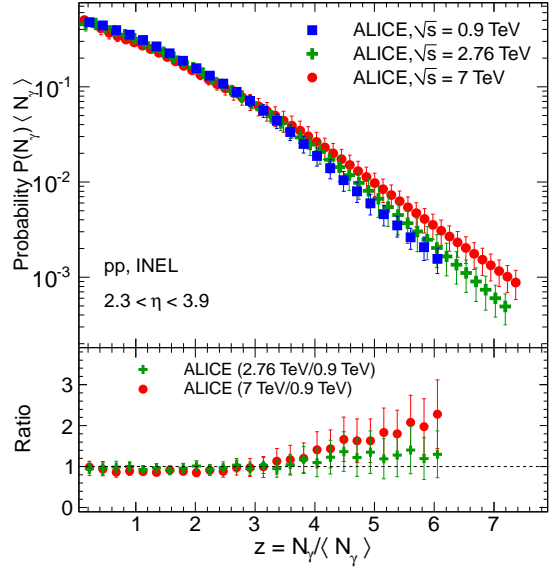}
\caption{
(Color online). Upper panel shows the photon multiplicity distributions in terms of KNO variable for pp
collisions at \sqrts~=~0.9, 2.76 and 7 TeV for inelastic collisions. 
Lower panel shows the ratios of the 
distributions for 2.76 and 7~TeV with respect to 0.9~TeV. 
}
\label{KNO}
\end{center}
\end{figure} 

\begin{figure}[tbh!]
\begin{center}
\includegraphics[width=0.5\textwidth,height=0.28\textheight]{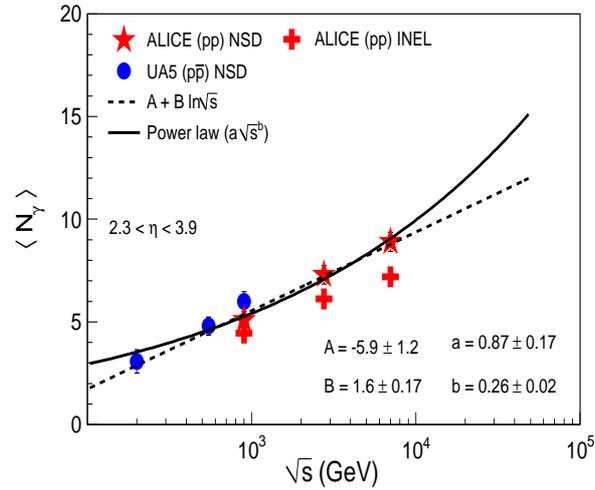}
\caption{(Color online).
Average photon multiplicity within \etawindow~as
a function of center of mass energy for \pp~collisions. The data
points from UA5 experiment~\cite{UA5} are superimposed on the ALICE
data. The energy dependence for the NSD events is consistent with both 
a logarithmic and a power-law fit.
}
\label{Energydep}
\end{center}
\end{figure}

Scaling property of multiplicity distributions of produced particles had been proposed in
1972 by Koba, Nielsen and Olesen (KNO)~\cite{KNO1}, based on the assumption of 
Feynman scaling of particle production~\cite{Feynman}. 
The multiplicity distributions, plotted in
terms of the KNO variable, $z = N_{\gamma}/\langle N_{\gamma}
\rangle$, are expected to be independent of collision energy~\cite{KNO1,KNO2}.
To test this scaling at the forward rapidities at the LHC energies,
the photon multiplicity distributions have been presented in terms of
the KNO variable.
The upper panel of Fig.~\ref{KNO} shows the distributions of
$z$ for the three energies, \sqrts~=~0.9, 2.76 and 7~TeV. 
The distributions are similar for low values of $z$, while for higher
$z$ values ($z>3$), the three distributions deviate from each other.
The deviations are studied by plotting the ratios of 
the $z$ values for \sqrts~=~2.76 and 7~TeV
with respect to \sqrts~=~0.9~TeV, shown in the lower panel of 
Fig.~\ref{KNO}. Because of the difference in the $z$ values, the
nearest values of $z$ are considered at other energies. 
The ratios are close to unity up to a value of $z=3$ and deviate from
unity for $z>3$.
Although the error bars at high multiplicities are large,
there is an indication of a mild deviation from the KNO scaling,
similar to what has been observed for charged particles at
mid-rapidity~\cite{ALICE-pp1}. Violations of KNO scaling for charged
particle multiplicities had been observed earlier for
pp and  ${\rm p}\bar{\rm p}$ collisions at
$\sqrt{s}$ = 30~GeV to 1800~GeV at the Fermilab Tevatron and also in UA5
experiment at CERN at 546~GeV~\cite{KNO3,KNO4,KNO5}. The present
observation at the LHC energies is consistent with these findings. 

\begin{figure}[tbh!]
\begin{center}
\includegraphics[width=0.5\textwidth]{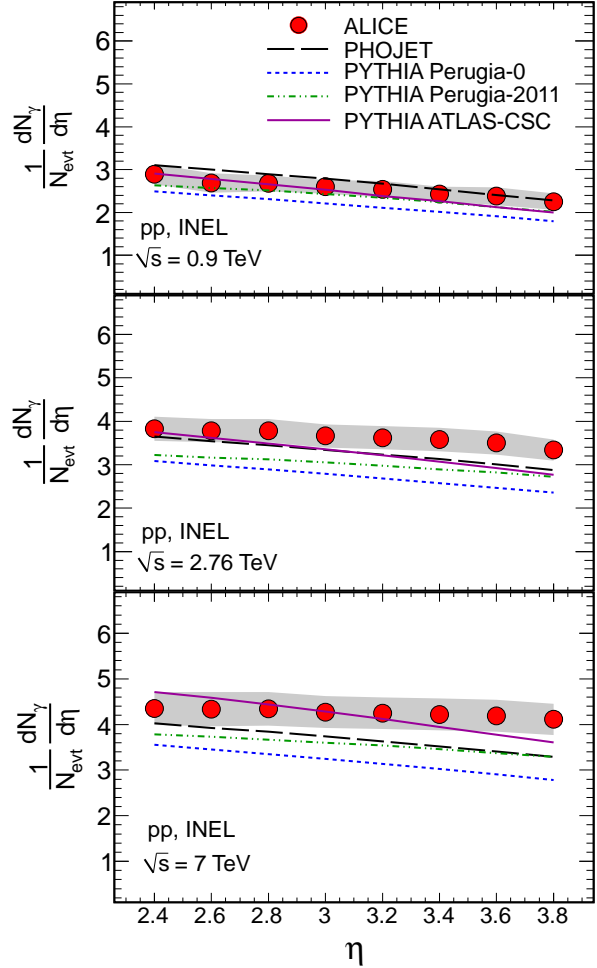}
\caption{(Color online). 
Pseudorapidity distribution of photons for inelastic events 
for \pp~collisions at \sqrts~=~0.9 (top), 2.76 (middle)
and 7~TeV (bottom). Results from different event generators 
are superimposed. The statistical errors are within the symbol sizes and 
the shaded regions represent the systematic uncertainties. 
}
\label{dNdEta}
\end{center}
\end{figure}

The energy dependences of the average photon multiplicity in INEL and
NSD events within \etawindow~are presented in 
Fig.~\ref{Energydep}. The NSD events are obtained using a special
trigger condition as discussed in Sec.~3. At lower center of mass
energies of 0.2, 0.546 and 0.9~TeV, the average photon multiplicities
for NSD events
have been measured by the UA5 experiment~\cite{UA5} within the same 
pseudorapidity region. It is seen that the
average photon multiplicity increases with increasing $\sqrt{s}$. The
nature of the increase has been studied in order to understand the
particle production mechanism and for extrapolating to higher energy
collisions. A logarithmic increase 
in average multiplicity with respect to $\sqrt{s}$ has been
predicted by the Feynman scaling of particle production~\cite{Feynman,reygers}.
For the NSD events, 
a logarithmic fit of the form ${\rm A}+{\rm B}~{\rm ln}\sqrt{s}$ yields
${\rm A} =-5.9\pm1.2$ and ${\rm B} =1.6\pm0.17$, where $\sqrt{s}$ is
expressed in GeV. 
On the other hand, phase space considerations give a power law
dependence of particle multiplicity~\cite{Berger}.
A power-law function of the form a$(\sqrt{s})^{\rm b}$ gives an equally
good fit to the NSD data points with ${\rm a} = 0.87\pm 0.17$ and
${\rm b} = 0.26\pm0.02$.
Thus, the present data cannot distinguish between the logarithmic and
power law dependence on $\sqrt{s}$.
At higher energies, the power-law fit predicts somewhat larger photon
multiplicity compared to the logarithmic function.
Data points at higher energies are needed to draw a conclusion on
the nature of the increase of photon multiplicity, and in this context
future LHC runs will be useful.

\begin{figure}[tbh!]
\begin{center}
\includegraphics[width=0.55\textwidth,height=0.28\textheight]{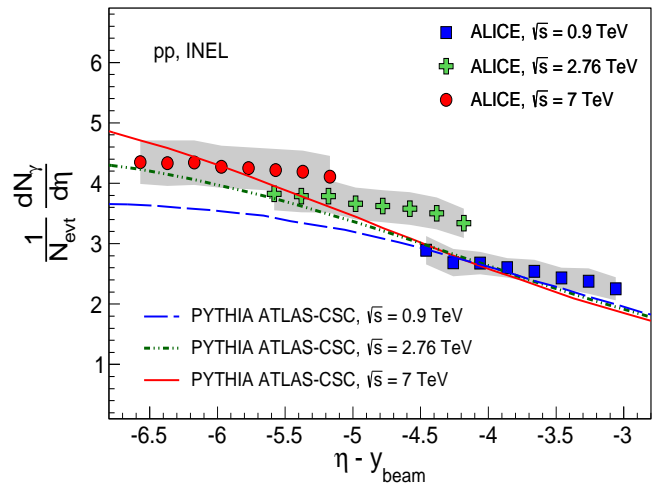}
\caption{(Color online). Pseudorapidity distribution of photons for
 inelastic events as a function of $\eta -  y_{\rm beam}$
for \pp~collisions at \sqrts~=~0.9, 2.76 and 7~TeV. Corresponding
distributions from PYTHIA~(tune ATLAS-CSC) are superimposed.
The statistical uncertainties are within symbol sizes and 
the shaded regions represent the systematic uncertainties. 
}
\label{LF}
\end{center}
\end{figure}

\subsection{Pseudorapidity distributions and limiting fragmentation}

Pseudorapidity distributions of photons for the INEL collisions have been obtained after
applying the unfolding method in each $\eta$~bin. In
Fig.~\ref{dNdEta}, the results for
pseudorapidity density are plotted as a function of pseudorapidity.
Results from PHOJET~1.2 and three different tunes of PYTHIA 6.4 
event generators are superimposed on the data points.
At $\sqrt{s} = 0.9$~TeV, all the calculations except 
PYTHIA (tune Perugia-0) describe the data within uncertainties,
whereas at $\sqrt{s} = 2.76$~TeV, PHOJET and PYTHIA (tune ATLAS-CSC) show a
better agreement compared to other event generators. 
At \sqrts~=~7~TeV, PYTHIA (tune ATLAS-CSC) are close to the data points,
whereas all other calculations under-predict the data. Thus 
PYTHIA (tune
ATLAS-CSC) results on the pseudorapidity distributions are the most
compatible with the measured data points, within the present uncertainties.

Particle productions at forward rapidity in \pp~and in heavy-ion
collisions are expected to follow the limiting fragmentation
behavior. This means that
particle production in the rest frame of one of the colliding hadrons
is independent of center-of-mass
energy~\cite{limiting1,limiting2,limiting3}. 
This phenomenon is attributed to the longitudinal scaling of particle
multiplicities. This behavior is studied by shifting the pseudorapidity bins
by the beam rapidity, $y_{\rm beam}$. 
The UA5 experiment at
CERN has observed the limiting fragmentation behavior in 
pp and  ${\rm p}\bar{\rm p}$ inelastic collisions from 53~GeV 
to 900~GeV~\cite{UA5_limiting} for charged particles in 
$|\eta - y_{\rm beam}| > -2.5$. Such observations have also been made 
for nucleus-nucleus collisions at RHIC energies~\cite{star,phobos,brahms}.
For the present analyses, the $y_{\rm beam}$ values
are 6.86, 7.98 and
8.97 at \sqrts~=~0.9, 2.76 and 7 TeV, respectively.
The results of
the pseudorapidity distributions of photons, after shifting for the beam
rapidity, are presented in Fig.~\ref{LF}.
The distributions from PYTHIA (tune ATLAS-CSC) event
generator are also presented in Fig.~\ref{LF} for all three energies, which
indicate that the limiting fragmentation behavior of photons hold for 
$|\eta - y_{\rm beam}| < -4.5$. 
In the range of the present measurement, 
limiting fragmentation behavior is not expected and not observed.
This suggests that at
the LHC this behavior may be confined to a  pseudorapidity
interval closer to beam rapidity.

\section{Summary}

We have measured the multiplicity and pseudorapidity distributions of photons
in \pp~collisions at $\sqrt{s} = 0.9, 2.76$ and 7~TeV in the forward
rapidity region (\etawindow)  using the PMD installed in the ALICE detector.
The results are
compared to the photon multiplicity distributions obtained with
different event generators.
It is observed that PHOJET explains the multiplicity distributions at
\sqrts~=~0.9~TeV, but under-predicts the data at
other energies. 
At \sqrts~=~2.76~TeV and 7~TeV,
the results from the ATLAS-CSC tune of PYTHIA are closer to the data
compared to all other event generators.
Photon multiplicity distributions are well described
by single NBD functions.
We observe deviations from KNO scaling for  $z>3$.
The energy dependence of the average photon multiplicity within
\etawindow~increases with the increase in \sqrts, and
is consistent with both a logarithmic and a
power-law dependence.
The pseudorapidity distributions of photons have been presented and compared
to the results from event generators. Results from
PYTHIA (tune ATLAS-CSC) are most
compatible with the measured data points compared to 
other generators. Longitudinal scaling of
photon production 
is not observed within the measured pseudorapidity range.
Future measurements at larger rapidities will help in better
understanding of the limiting fragmentation behavior.
%%%%%%%%%%% put the body of the article here
% 

%%%%% acknowledgements
\newenvironment{acknowledgement}{\relax}{\relax}
\begin{acknowledgement}
\section*{Acknowledgements}
The ALICE Collaboration would like to thank all its engineers and technicians for their invaluable contributions to the construction of the experiment and the CERN accelerator teams for the outstanding performance of the LHC complex.
%\\
The ALICE Collaboration gratefully acknowledges the resources and support provided by all Grid centres and the Worldwide LHC Computing Grid (WLCG) collaboration.
%\\
The ALICE Collaboration acknowledges the following funding agencies for their support in building and
running the ALICE detector:
 %\\
State Committee of Science,  World Federation of Scientists (WFS)
and Swiss Fonds Kidagan, Armenia,
 %\\
Conselho Nacional de Desenvolvimento Cient\'{\i}fico e Tecnol\'{o}gico (CNPq), Financiadora de Estudos e Projetos (FINEP),
Funda\c{c}\~{a}o de Amparo \`{a} Pesquisa do Estado de S\~{a}o Paulo (FAPESP);
 %\\
National Natural Science Foundation of China (NSFC), the Chinese Ministry of Education (CMOE)
and the Ministry of Science and Technology of China (MSTC);
 %\\
Ministry of Education and Youth of the Czech Republic;
 %\\
Danish Natural Science Research Council, the Carlsberg Foundation and the Danish National Research Foundation;
 %\\
The European Research Council under the European Community's Seventh Framework Programme;
 %\\
Helsinki Institute of Physics and the Academy of Finland;
 %\\
French CNRS-IN2P3, the `Region Pays de Loire', `Region Alsace', `Region Auvergne' and CEA, France;
 %\\
German BMBF and the Helmholtz Association;
%\\
General Secretariat for Research and Technology, Ministry of
Development, Greece;
%\\
Hungarian OTKA and National Office for Research and Technology (NKTH);
 %\\
Department of Atomic Energy and Department of Science and Technology of the Government of India;
 %\\
Istituto Nazionale di Fisica Nucleare (INFN) and Centro Fermi -
Museo Storico della Fisica e Centro Studi e Ricerche "Enrico
Fermi", Italy;
 %\\
MEXT Grant-in-Aid for Specially Promoted Research, Ja\-pan;
 %\\
Joint Institute for Nuclear Research, Dubna;
 %\\
%Korea Foundation for International Cooperation of Science and Technology (KICOS);
National Research Foundation of Korea (NRF);
 %\\
CONACYT, DGAPA, M\'{e}xico, ALFA-EC and the EPLANET Program
(European Particle Physics Latin American Network)
 %\\
Stichting voor Fundamenteel Onderzoek der Materie (FOM) and the Nederlandse Organisatie voor Wetenschappelijk Onderzoek (NWO), Netherlands;
 %\\
Research Council of Norway (NFR);
 %\\
Polish Ministry of Science and Higher Education;
 %\\
National Science Centre, Poland;
 %\\
 Ministry of National Education/Institute for Atomic Physics and CNCS-UEFISCDI - Romania;
 %\\
Ministry of Education and Science of Russian Federation, Russian
Academy of Sciences, Russian Federal Agency of Atomic Energy,
Russian Federal Agency for Science and Innovations and The Russian
Foundation for Basic Research;
 %\\
Ministry of Education of Slovakia;
 %\\
Department of Science and Technology, South Africa;
 %\\
CIEMAT, EELA, Ministerio de Econom\'{i}a y Competitividad (MINECO) of Spain, Xunta de Galicia (Conseller\'{\i}a de Educaci\'{o}n),
CEA\-DEN, Cubaenerg\'{\i}a, Cuba, and IAEA (International Atomic Energy Agency);
 %\\
Swedish Research Council (VR) and Knut $\&$ Alice Wallenberg
Foundation (KAW);
 %\\
Ukraine Ministry of Education and Science;
 %\\
United Kingdom Science and Technology Facilities Council (STFC);
 %\\
The United States Department of Energy, the United States National
Science Foundation, the State of Texas, and the State of Ohio;
%\\
Ministry of Science, Education and Sports of Croatia and  Unity through Knowledge Fund, Croatia.

    %%%%%%% done by webmaster team
\end{acknowledgement}

%%%%%%%% Bibliography (In case of using bibtex generate the bbl requested by arXiv)
%\bibliographystyle{utphys_notitle}   % Put here the style file name
%for the paper, i.e.apsrev4-1

\bibliographystyle{utphys}   % Put here the style file name for the paper, i.e.apsrev4-1
\bibliography{biblio_ppPhoton}

\newpage
% \input{}
% 
%%%%%%%%% appendix with author list
\appendix
\section{ALICE Collaboration}
\label{app:collab}
\label{app:collab}

% Collaboration: CERN-LHC-ALICE
% Generation Date is 2014/Jun/18

% How to use:
%%%%%%%%% appendix with author list
%\appendix
%\section{The ALICE Collaboration}
%\label{app:collab}
%\input{authors-list.tex}  %%%%%%% get the latest version before submitting

\begingroup
\small
\begin{flushleft}
B.~Abelev\Irefn{org71}\And
J.~Adam\Irefn{org37}\And
D.~Adamov\'{a}\Irefn{org79}\And
M.M.~Aggarwal\Irefn{org83}\And
G.~Aglieri~Rinella\Irefn{org34}\And
M.~Agnello\Irefn{org107}\textsuperscript{,}\Irefn{org90}\And
A.~Agostinelli\Irefn{org26}\And
N.~Agrawal\Irefn{org44}\And
Z.~Ahammed\Irefn{org126}\And
N.~Ahmad\Irefn{org18}
% ...
\renewcommand\labelenumi{\textsuperscript{\theenumi}~}

\section*{Affiliation notes}
\renewcommand\theenumi{\roman{enumi}}
\begin{Authlist}
\item \Adef{0}Deceased
\item \Adef{idp1126816}{Also at: St. Petersburg State Polytechnical University}
\item \Adef{idp3061824}{Also at: Department of Applied Physics, Aligarh Muslim University, Aligarh, India}
\item \Adef{idp3741600}{Also at: M.V. Lomonosov Moscow State University, D.V. Skobeltsyn Institute of Nuclear Physics, Moscow, Russia}
\item \Adef{idp3998400}{Also at: University of Belgrade, Faculty of Physics and "Vin\v{c}a" Institute of Nuclear Sciences, Belgrade, Serbia}
\item \Adef{idp4317104}{Permanent Address: Permanent Address: Konkuk University, Seoul, Korea}
\item \Adef{idp4880768}{Also at: Institute of Theoretical Physics, University of Wroclaw, Wroclaw, Poland}
\item \Adef{idp5813632}{Also at: University of Kansas, Lawrence, KS, United States}
\end{Authlist}

\section*{Collaboration Institutes}
\renewcommand\theenumi{\arabic{enumi}~}
\begin{Authlist}

\item \Idef{org1}A.I. Alikhanyan National Science Laboratory (Yerevan Physics Institute) Foundation, Yerevan, Armenia
\item \Idef{org2}Benem\'{e}rita Universidad Aut\'{o}noma de Puebla, Puebla, Mexico
\item \Idef{org3}Bogolyubov Institute for Theoretical Physics, Kiev, Ukraine
\item \Idef{org4}Bose Institute, Department of Physics and Centre for Astroparticle Physics and Space Science (CAPSS), Kolkata, India
\item \Idef{org5}Budker Institute for Nuclear Physics, Novosibirsk, Russia
\item \Idef{org6}California Polytechnic State University, San Luis Obispo, CA, United States
\item \Idef{org7}Central China Normal University, Wuhan, China
\item \Idef{org8}Centre de Calcul de l'IN2P3, Villeurbanne, France
\item \Idef{org9}Centro de Aplicaciones Tecnol\'{o}gicas y Desarrollo Nuclear (CEADEN), Havana, Cuba
\item \Idef{org10}Centro de Investigaciones Energ\'{e}ticas Medioambientales y Tecnol\'{o}gicas (CIEMAT), Madrid, Spain
\item \Idef{org11}Centro de Investigaci\'{o}n y de Estudios Avanzados (CINVESTAV), Mexico City and M\'{e}rida, Mexico
\item \Idef{org12}Centro Fermi - Museo Storico della Fisica e Centro Studi e Ricerche ``Enrico Fermi'', Rome, Italy
\item \Idef{org13}Chicago State University, Chicago, USA
\item \Idef{org14}Commissariat \`{a} l'Energie Atomique, IRFU, Saclay, France
\item \Idef{org15}COMSATS Institute of Information Technology (CIIT), Islamabad, Pakistan
\item \Idef{org16}Departamento de F\'{\i}sica de Part\'{\i}culas and IGFAE, Universidad de Santiago de Compostela, Santiago de Compostela, Spain
\item \Idef{org17}Department of Physics and Technology, University of Bergen, Bergen, Norway
\item \Idef{org18}Department of Physics, Aligarh Muslim University, Aligarh, India
\item \Idef{org19}Department of Physics, Ohio State University, Columbus, OH, United States
\item \Idef{org20}Department of Physics, Sejong University, Seoul, South Korea
\item \Idef{org21}Department of Physics, University of Oslo, Oslo, Norway
\item \Idef{org22}Dipartimento di Fisica dell'Universit\`{a} 'La Sapienza' and Sezione INFN Rome, Italy
\item \Idef{org23}Dipartimento di Fisica dell'Universit\`{a} and Sezione INFN, Cagliari, Italy
\item \Idef{org24}Dipartimento di Fisica dell'Universit\`{a} and Sezione INFN, Trieste, Italy
\item \Idef{org25}Dipartimento di Fisica dell'Universit\`{a} and Sezione INFN, Turin, Italy
\item \Idef{org26}Dipartimento di Fisica e Astronomia dell'Universit\`{a} and Sezione INFN, Bologna, Italy
\item \Idef{org27}Dipartimento di Fisica e Astronomia dell'Universit\`{a} and Sezione INFN, Catania, Italy
\item \Idef{org28}Dipartimento di Fisica e Astronomia dell'Universit\`{a} and Sezione INFN, Padova, Italy
\item \Idef{org29}Dipartimento di Fisica `E.R.~Caianiello' dell'Universit\`{a} and Gruppo Collegato INFN, Salerno, Italy
\item \Idef{org30}Dipartimento di Scienze e Innovazione Tecnologica dell'Universit\`{a} del  Piemonte Orientale and Gruppo Collegato INFN, Alessandria, Italy
\item \Idef{org31}Dipartimento Interateneo di Fisica `M.~Merlin' and Sezione INFN, Bari, Italy
\item \Idef{org32}Division of Experimental High Energy Physics, University of Lund, Lund, Sweden
\item \Idef{org33}Eberhard Karls Universit\"{a}t T\"{u}bingen, T\"{u}bingen, Germany
\item \Idef{org34}European Organization for Nuclear Research (CERN), Geneva, Switzerland
\item \Idef{org35}Faculty of Engineering, Bergen University College, Bergen, Norway
\item \Idef{org36}Faculty of Mathematics, Physics and Informatics, Comenius University, Bratislava, Slovakia
\item \Idef{org37}Faculty of Nuclear Sciences and Physical Engineering, Czech Technical University in Prague, Prague, Czech Republic
\item \Idef{org38}Faculty of Science, P.J.~\v{S}af\'{a}rik University, Ko\v{s}ice, Slovakia
\item \Idef{org39}Frankfurt Institute for Advanced Studies, Johann Wolfgang Goethe-Universit\"{a}t Frankfurt, Frankfurt, Germany
\item \Idef{org40}Gangneung-Wonju National University, Gangneung, South Korea
\item \Idef{org41}Gauhati University, Department of Physics, Guwahati, India
\item \Idef{org42}Helsinki Institute of Physics (HIP), Helsinki, Finland
\item \Idef{org43}Hiroshima University, Hiroshima, Japan
\item \Idef{org44}Indian Institute of Technology Bombay (IIT), Mumbai, India
\item \Idef{org45}Indian Institute of Technology Indore, Indore (IITI), India
\item \Idef{org46}Inha University, Incheon, South Korea
\item \Idef{org47}Institut de Physique Nucl\'eaire d'Orsay (IPNO), Universit\'e Paris-Sud, CNRS-IN2P3, Orsay, France
\item \Idef{org48}Institut f\"{u}r Informatik, Johann Wolfgang Goethe-Universit\"{a}t Frankfurt, Frankfurt, Germany
\item \Idef{org49}Institut f\"{u}r Kernphysik, Johann Wolfgang Goethe-Universit\"{a}t Frankfurt, Frankfurt, Germany
\item \Idef{org50}Institut f\"{u}r Kernphysik, Westf\"{a}lische Wilhelms-Universit\"{a}t M\"{u}nster, M\"{u}nster, Germany
\item \Idef{org51}Institut Pluridisciplinaire Hubert Curien (IPHC), Universit\'{e} de Strasbourg, CNRS-IN2P3, Strasbourg, France
\item \Idef{org52}Institute for Nuclear Research, Academy of Sciences, Moscow, Russia
\item \Idef{org53}Institute for Subatomic Physics of Utrecht University, Utrecht, Netherlands
\item \Idef{org54}Institute for Theoretical and Experimental Physics, Moscow, Russia
\item \Idef{org55}Institute of Experimental Physics, Slovak Academy of Sciences, Ko\v{s}ice, Slovakia
\item \Idef{org56}Institute of Physics, Academy of Sciences of the Czech Republic, Prague, Czech Republic
\item \Idef{org57}Institute of Physics, Bhubaneswar, India
\item \Idef{org58}Institute of Space Science (ISS), Bucharest, Romania
\item \Idef{org59}Instituto de Ciencias Nucleares, Universidad Nacional Aut\'{o}noma de M\'{e}xico, Mexico City, Mexico
\item \Idef{org60}Instituto de F\'{\i}sica, Universidad Nacional Aut\'{o}noma de M\'{e}xico, Mexico City, Mexico
\item \Idef{org61}iThemba LABS, National Research Foundation, Somerset West, South Africa
\item \Idef{org62}Joint Institute for Nuclear Research (JINR), Dubna, Russia
\item \Idef{org63}Konkuk University, Seoul, South Korea
\item \Idef{org64}Korea Institute of Science and Technology Information, Daejeon, South Korea
\item \Idef{org65}KTO Karatay University, Konya, Turkey
\item \Idef{org66}Laboratoire de Physique Corpusculaire (LPC), Clermont Universit\'{e}, Universit\'{e} Blaise Pascal, CNRS--IN2P3, Clermont-Ferrand, France
\item \Idef{org67}Laboratoire de Physique Subatomique et de Cosmologie, Universit\'{e} Grenoble-Alpes, CNRS-IN2P3, Grenoble, France
\item \Idef{org68}Laboratori Nazionali di Frascati, INFN, Frascati, Italy
\item \Idef{org69}Laboratori Nazionali di Legnaro, INFN, Legnaro, Italy
\item \Idef{org70}Lawrence Berkeley National Laboratory, Berkeley, CA, United States
\item \Idef{org71}Lawrence Livermore National Laboratory, Livermore, CA, United States
\item \Idef{org72}Moscow Engineering Physics Institute, Moscow, Russia
\item \Idef{org73}National Centre for Nuclear Studies, Warsaw, Poland
\item \Idef{org74}National Institute for Physics and Nuclear Engineering, Bucharest, Romania
\item \Idef{org75}National Institute of Science Education and Research, Bhubaneswar, India
\item \Idef{org76}Niels Bohr Institute, University of Copenhagen, Copenhagen, Denmark
\item \Idef{org77}Nikhef, National Institute for Subatomic Physics, Amsterdam, Netherlands
\item \Idef{org78}Nuclear Physics Group, STFC Daresbury Laboratory, Daresbury, United Kingdom
\item \Idef{org79}Nuclear Physics Institute, Academy of Sciences of the Czech Republic, \v{R}e\v{z} u Prahy, Czech Republic
\item \Idef{org80}Oak Ridge National Laboratory, Oak Ridge, TN, United States
\item \Idef{org81}Petersburg Nuclear Physics Institute, Gatchina, Russia
\item \Idef{org82}Physics Department, Creighton University, Omaha, NE, United States
\item \Idef{org83}Physics Department, Panjab University, Chandigarh, India
\item \Idef{org84}Physics Department, University of Athens, Athens, Greece
\item \Idef{org85}Physics Department, University of Cape Town, Cape Town, South Africa
\item \Idef{org86}Physics Department, University of Jammu, Jammu, India
\item \Idef{org87}Physics Department, University of Rajasthan, Jaipur, India
\item \Idef{org88}Physik Department, Technische Universit\"{a}t M\"{u}nchen, Munich, Germany
\item \Idef{org89}Physikalisches Institut, Ruprecht-Karls-Universit\"{a}t Heidelberg, Heidelberg, Germany
\item \Idef{org90}Politecnico di Torino, Turin, Italy
\item \Idef{org91}Purdue University, West Lafayette, IN, United States
\item \Idef{org92}Pusan National University, Pusan, South Korea
\item \Idef{org93}Research Division and ExtreMe Matter Institute EMMI, GSI Helmholtzzentrum f\"ur Schwerionenforschung, Darmstadt, Germany
\item \Idef{org94}Rudjer Bo\v{s}kovi\'{c} Institute, Zagreb, Croatia
\item \Idef{org95}Russian Federal Nuclear Center (VNIIEF), Sarov, Russia
\item \Idef{org96}Russian Research Centre Kurchatov Institute, Moscow, Russia
\item \Idef{org97}Saha Institute of Nuclear Physics, Kolkata, India
\item \Idef{org98}School of Physics and Astronomy, University of Birmingham, Birmingham, United Kingdom
\item \Idef{org99}Secci\'{o}n F\'{\i}sica, Departamento de Ciencias, Pontificia Universidad Cat\'{o}lica del Per\'{u}, Lima, Peru
\item \Idef{org100}Sezione INFN, Bari, Italy
\item \Idef{org101}Sezione INFN, Bologna, Italy
\item \Idef{org102}Sezione INFN, Cagliari, Italy
\item \Idef{org103}Sezione INFN, Catania, Italy
\item \Idef{org104}Sezione INFN, Padova, Italy
\item \Idef{org105}Sezione INFN, Rome, Italy
\item \Idef{org106}Sezione INFN, Trieste, Italy
\item \Idef{org107}Sezione INFN, Turin, Italy
\item \Idef{org108}SSC IHEP of NRC Kurchatov institute, Protvino, Russia
\item \Idef{org109}SUBATECH, Ecole des Mines de Nantes, Universit\'{e} de Nantes, CNRS-IN2P3, Nantes, France
\item \Idef{org110}Suranaree University of Technology, Nakhon Ratchasima, Thailand
\item \Idef{org111}Technical University of Split FESB, Split, Croatia
\item \Idef{org112}The Henryk Niewodniczanski Institute of Nuclear Physics, Polish Academy of Sciences, Cracow, Poland
\item \Idef{org113}The University of Texas at Austin, Physics Department, Austin, TX, USA
\item \Idef{org114}Universidad Aut\'{o}noma de Sinaloa, Culiac\'{a}n, Mexico
\item \Idef{org115}Universidade de S\~{a}o Paulo (USP), S\~{a}o Paulo, Brazil
\item \Idef{org116}Universidade Estadual de Campinas (UNICAMP), Campinas, Brazil
\item \Idef{org117}University of Houston, Houston, TX, United States
\item \Idef{org118}University of Jyv\"{a}skyl\"{a}, Jyv\"{a}skyl\"{a}, Finland
\item \Idef{org119}University of Liverpool, Liverpool, United Kingdom
\item \Idef{org120}University of Tennessee, Knoxville, TN, United States
\item \Idef{org121}University of Tokyo, Tokyo, Japan
\item \Idef{org122}University of Tsukuba, Tsukuba, Japan
\item \Idef{org123}University of Zagreb, Zagreb, Croatia
\item \Idef{org124}Universit\'{e} de Lyon, Universit\'{e} Lyon 1, CNRS/IN2P3, IPN-Lyon, Villeurbanne, France
\item \Idef{org125}V.~Fock Institute for Physics, St. Petersburg State University, St. Petersburg, Russia
\item \Idef{org126}Variable Energy Cyclotron Centre, Kolkata, India
\item \Idef{org127}Vestfold University College, Tonsberg, Norway
\item \Idef{org128}Warsaw University of Technology, Warsaw, Poland
\item \Idef{org129}Wayne State University, Detroit, MI, United States
\item \Idef{org130}Wigner Research Centre for Physics, Hungarian Academy of Sciences, Budapest, Hungary
\item \Idef{org131}Yale University, New Haven, CT, United States
\item \Idef{org132}Yonsei University, Seoul, South Korea
\item \Idef{org133}Zentrum f\"{u}r Technologietransfer und Telekommunikation (ZTT), Fachhochschule Worms, Worms, Germany
\end{Authlist}
\endgroup

  %%%%%%% done by webmaster team
\end{document}